\documentclass[sigconf]{acmart}

\AtBeginDocument{%
  }

\copyrightyear{2024} 
\acmYear{2024} 
\setcopyright{rightsretained} 
\acmConference[CHI '24]{Proceedings of the CHI Conference on Human Factors in Computing Systems}{May 11--16, 2024}{Honolulu, HI, USA}
\acmBooktitle{Proceedings of the CHI Conference on Human Factors in Computing Systems (CHI '24), May 11--16, 2024, Honolulu, HI, USA}
\acmDOI{10.1145/3613904.3642433}
\acmISBN{979-8-4007-0330-0/24/05}

\newcommand{\descr}[1]{ \noindent \textbf{#1}}

\usepackage{xcolor}
\usepackage{subfigure}
\usepackage{multirow}
\usepackage{ifthen}

\usepackage{enumitem}

\newif\ifrevision
\revisionfalse
\ifrevision
\newcommand{\revision}[1]{\textcolor{blue}{#1}}
\newcommand{\revisioncomment}[1]{\textbf{\textcolor{red}{[#1]}}}
\else
\newcommand{\revision}[1]{\textcolor{black}{#1}}
\newcommand{\revisioncomment}[1]{}
\fi

\begin{document}
\title[Analyzing User Engagement with TikTok's Short Format Video Recommendations using Data Donations]{Analyzing User Engagement with TikTok's Short Format Video Recommendations using Data Donations}

\author{Savvas Zannettou}
\affiliation{%
  \institution{TU Delft}
  \city{Delft}
  \country{The Netherlands}
}

\affiliation{%
  \institution{Max Planck Institute for Informatics}
  \city{Saarbrücken}
  \country{Germany}
}
\author{Olivia Nemes-Nemeth}
\affiliation{%
  \institution{Max Planck Institute for Software Systems}
  \city{Saarbrücken}
  \country{Germany}
}
\author{Oshrat Ayalon}
\affiliation{%
  \institution{University of Haifa}
  \city{Haifa}
  \country{Israel}
}
\affiliation{%
  \institution{Max Planck Institute for Software Systems}
  \city{Saarbrücken}
  \country{Germany}
}
\author{Angelica Goetzen}
\affiliation{%
  \institution{Max Planck Institute for Software Systems}
  \city{Saarbrücken}
  \country{Germany}
}

\author{Krishna P. Gummadi}
\affiliation{%
  \institution{Max Planck Institute for Software Systems}
  \city{Saarbrücken}
  \country{Germany}
}

\author{Elissa M. Redmiles}
\authornote{Completed a portion of this work while at the Max Planck Institute for Software Systems.}
\affiliation{%
  \institution{Georgetown University}
  \city{Washington D.C.}
  \country{USA}
}

\author{Franziska Roesner}
\affiliation{%
  \institution{University of Washington}
  \city{Seattle}
  \country{USA}
}
\renewcommand{\shortauthors}{Zannettou et al.}

\begin{abstract}
Short-format videos have exploded on platforms like TikTok, Instagram, and YouTube.
Despite this, the research community lacks large-scale empirical studies into how people engage with short-format videos and the role of recommendation systems that offer endless streams of such content. 
In this work, we analyze user engagement on TikTok using data we collect via a data donation system that allows TikTok users to donate their data.
We recruited 347 TikTok users and collected 9.2M TikTok video recommendations they received.
By analyzing user engagement, we find that the average daily usage time increases over the users' lifetime while the user attention remains stable at around 45\%.
We also find that users like more videos uploaded by people they follow than those recommended by people they do not follow.
Our study offers valuable insights into how users engage with short-format videos on TikTok and lessons learned from designing a data donation system.

\end{abstract}


\maketitle

\section{Introduction}

We are witnessing a significant shift in how people consume and engage with social media content.
This shift is the result of two new trends on social media platforms: first, the rise in popularity of {\emph short-format videos} (i.e., videos that are less than 60 seconds), and second, the rise of algorithmic recommendation systems that offer {\emph endless streams of personalized recommendations} of these videos, requiring no explicit inputs from users.
A social media platform that pioneered and exemplified these trends is TikTok, a conglomeration of traditional social networking features, content shared via short-format videos, and a recommendation algorithm that offers never-ending streams of video recommendations. 
TikTok is widely popular, with more than 1.3 billion users worldwide~\cite{businessofapps1}.
In response to TikTok's popularity, other platforms like Facebook (Reels~\cite{facebook_reels}), YouTube (Shorts~\cite{youtube_shorts}), Instagram (Reels~\cite{instagram_reels}), and even Netflix (Fast Laughs~\cite{netflix_laughs}) started offering their own short-format video feature, powered by recommendation algorithms.

Despite the rapidly growing adoption and use of this short-format video feature powered by algorithmic recommendations, we lack empirical studies on how users engage with short-format videos and what role the recommendation algorithm plays in users' content consumption online.
Anecdotal evidence and journalistic investigative reports~\cite{nyt_how,wsj_investigation} highlight that TikTok's recommendation algorithm is very effective and can accurately recommend videos that users find interesting or engaging. 
At the same time, there are some concerns that the recommendation algorithm may lead users towards problematic content~\cite{wsj_investigation}, which emphasizes the need to rigorously audit the effectiveness of recommendation algorithms and their role in users' content consumption/engagement.
Overall, beyond these anecdotal evidence and journalistic efforts based on traces from automated accounts, there are a few empirical studies (e.g.,~\cite{boeker2022empirical}) that rigorously examine the role and effectiveness of algorithms that power short-format video recommendations.

In this paper, we attempt to bridge this research gap by focusing on the effectiveness of TikTok's recommendation algorithm by analyzing users' engagement with content recommendations.
To do this, we implement a data donation system that enables us to obtain real traces from TikTok users that leverage their right of access to data subjects as documented in EU's General Data Protection Regulation~\cite{gdpr}.
We recruited 347 real TikTok users and obtained their viewing history and associated engagement signals, including 9.2M video views.
Then, we analyze user engagement with short-format videos to shed light on the effectiveness of TikTok's recommendation algorithm, using empirical and authentic traces from real users.
We use three signals to measure the effectiveness of the TikTok algorithm through the lens of user engagement:
1)~Total amount of time spent and volume of videos consumed by TikTok users;
2)~Whether TikTok users watch videos till the end (\emph{Attention});
3)~Whether they interact with recommended videos (e.g., by liking videos).
Our analysis focuses on understanding the prevalence of the above-mentioned signals and, more importantly, how these signals change over time, which allows us to extract insights into the effectiveness of TikTok's recommendations.
\revision{Based on previous work~\cite{zou2019reinforcement}, an effective recommendation system should be able to maximize the user's interaction with the social feed in terms of engagement (e.g., liking recommendations or staying on TikTok for longer periods).
In this work, we combine the user engagement objectives of recommendation systems, methods to measure user engagement by~\cite{lalmas2022measuring}, as well as platform-specific affordances from TikTok, and we create the following constructs to measure the effectiveness of TikTok's recommendation algorithm through the lens of user engagement:
\begin{itemize}
    \item An effective recommendation system should increase the number of videos a user watches until the end and the number of liked videos.
    \item An effective recommendation system should increase the number of videos watched until the end and liked over time. Also, it should increase the time spent on the platform and the number of videos watched over time. \revisioncomment{R2}
\end{itemize}
}

Our analysis focuses on studying the above-mentioned constructs. 
Additionally, we complement our analysis by investigating differences in user engagement for videos originating from users' social networks (i.e., from accounts that users follow) or not, as this is a signal that provides explicit feedback about a user's interests to the recommendation algorithm.

\descr{Main Findings.} The main findings from our analysis are:
\begin{itemize}
\item Over time, the TikTok participants' daily average number of videos viewed and time spent on the platform is increasing ($2\times$ increase after 80 days).
\item Across all the participants, we find that 55\% of the recommended videos were not watched till the end (with most skipped before reaching the halfway point). When analyzing the temporal dimension, we find that, over time, the percentage of videos watched till the end is stable.
\item Over time, our participants' interaction with the recommended videos via the liking feature is increasing ($2\times$ increase after 120 days for videos from following and $1.5\times$ increase after 120 days for videos from non-following accounts).
\item TikTok users tend to pay more attention (i.e., watch until the end) to recommended videos from accounts they don't follow than those they do. Videos from non-following accounts are significantly more popular on TikTok than videos from following accounts. This may explain why the participants watched more of those videos until the end and why TikTok is likely limiting the recommendations from following accounts. 
\end{itemize}

\descr{Contributions.} We make the following notable contributions:
\begin{itemize}
    \item \revision{We perform a large-scale empirical analysis of how users engage with short-format videos on TikTok. In contrast with previous work (e.g.,~\cite{wsj_investigation,boeker2022empirical}), we use data from real users that are more authentic and diverse than traces obtained from automated accounts.}\revisioncomment{R2}
    \item \revision{We shed some light on the effectiveness of TikTok's recommendation algorithm through the lens of user engagement.
    We show that the volume of videos/time spent on the platform and user engagement through liking increases over time.
    At the same time, we find that the attention from the TikTok participants does not increase over time and that most participants watch between 30\% and 50\% of all videos till the end. 
    These empirical insights likely indicate that TikTok's algorithm prioritizes increasing the time spent on the platform and user engagement (through liking) rather than making recommendations that are likely to be watched until the end by users.
    Overall, this empirical evidence emphasizes the need to analyze these recommendation algorithms further and improve our understanding of the interplay between recommendation algorithms and user engagement on short-format video platforms. }\revisioncomment{R2}
    \item We demonstrate how data donation can be used to perform empirical studies on social media platforms. We argue that this is a promising avenue for future work aiming to collect and analyze behavioral traces from real users rather than automated accounts/profiles implemented by researchers. Also, we share lessons learned from conducting our data collection and future avenues for obtaining real traces via data donation.
    Given recent changes to data access by social media platforms like Twitter and Reddit that essentially preclude access to large-scale datasets to researchers~\cite{twitter_api_rip,reddit_api_changes}, we believe that collecting datasets using data donations and citizen science is an effective alternative way to perform research studies with data that is otherwise hard to obtain.
    Our experience and lessons learned from our data collection efforts will be invaluable to the research community and future endeavors that aim to collect large-scale datasets via data donations.
    \revision{The proposed approach is versatile and can be applied for obtaining data from any social media platform that allows users to obtain their data after GDPR data access requests.}
\end{itemize}

\section{Background \& Related Work}

\subsection{TikTok}
TikTok is a short-format video platform launched in 2016~\cite{tiktokbbc2,sun2020content,kaye2022tiktok}. 
Since then, TikTok has become the most downloaded app of 2020~\cite{tiktokbbc1}. 
As a platform, TikTok allows users to both watch and create short-format videos up to 10 minutes in length~\cite{techcrunch1}. TikTok offers multiple editing capabilities for creators and allows users to connect with peers via following, messaging, and sharing content. One of the app's most prominent characteristics is its ability to recommend relevant video content to viewers \cite{nyt_how,klug2021trick}; when using TikTok, a user may scroll through two different content feeds, one containing videos posted by the people they follow (“Following”), the other a curated feed of content from many different creators (“For You”).
Much of the prior work on TikTok has studied the recommendation algorithm, the content on the platform, and the users.

\descr{Recommendation Algorithm.} 
TikTok itself has stated that user interactions, video information, and device/account information impact a user's For You feed~\cite{tiktok_2020}, with no further information on what other data may play a role or how much each factor matters. Research efforts are being made to understand recommendations via algorithm audits. At present, it is difficult to assess whether a video is popularized by user engagement or by systematic amplification from the algorithm~\cite{bandy2020tulsaflop}. Sock-puppet audits have found evidence that a user's language, location, use of the ``follow'' and ``like'' features, and video viewing length all impact the contents of a user's For You feed, with the use of the ``follow'' feature exerting the strongest influence~\cite{boeker2022empirical}. ``Time of posting'' has also been identified as a relevant factor~\cite{klug2021trick}. Journalistic investigations have discussed the dangers of the algorithm, including its potential to drag users down rabbit holes of harmful content~\cite{wsj_investigation,nyt_how}. 

\descr{Content Analyses.} Prior work has also surveyed the content posted on TikTok like sentiments towards particular topics (e.g.,~\cite{basch2022climate,fowler2021sex,purushothaman2022content,yeung2022tiktok}), the proliferation of viral trends and memes~\cite{ling2021slapping,zeng2021okboomer}, and differences in content between TikTok and other platforms \cite{sun2020content,literat2021popular}. Researchers also analyze TikTok content to uncover correlations between types of content and user engagement~\cite{basch2021global,chen2021factors,ling2022learn}. Content analyses have additionally been used to audit the quality of information on the app~\cite{song2021short} and the presence of harmful content, with research finding that extremist, far-right, and anti-Semitic content has a robust presence~\cite{weimann2020research, weimann2021tiktok}. Inaccuracies in TikTok's use of warning labels have been discovered via content analysis~\cite{ling2022learn}, showcasing that detecting misinformation on a platform that affords concurrent audio, video, and text content can be challenging~\cite{shang2021multimodal}. 

\descr{User Studies.} Users have identified best practices for interacting with their For You feed in order to personalize its content, as well as formed beliefs on how the algorithm may systematically suppress or uplift certain content \cite{zeng2022content,simpson2021you,karizat2021algorithmic}. Video engagement, posting time, and piling hashtags stand out to users as driving factors for virality~\cite{klug2021trick}. 
In studying motivations to use the platform, researchers find escapism, social interaction, and archiving videos are primary motivators for passive TikTok use among U.S. users, while self-expression motivates participation on the platform~\cite{omar2020watch}; entertainment drives of all types of TikTok use in Denmark \cite{bossen2020uses}; and entertainment, learning new information, socially-rewarding self-expression, trendiness, escapism, and novelty are drivers in China~\cite{scherr2021explaining,lu2020exploring}. Motivations for non-use are also studied; among Chinese users who leave the platform, users cite fears of addiction or perceiving the content as low quality~\cite{lu2020exploring}, or requiring temporary focus on other tasks as motivations for non-use~\cite{lu2020exploring}. In terms of TikTok's potential negative impacts, neither passive nor active use of TikTok was found to relate to individuals' well-being during the COVID-19 pandemic~\cite{masciantonio2021don}. TikTok's impact on marginalized communities like LGBTQ+ identified individuals are also being investigated~\cite{simpson2021you}. With significant app engagement and a growing user base, the impact of TikTok on its users continues to be an important area of study.

\subsection{User Engagement}
\revision{Prior work focused on operationalizing, understanding, and analyzing user engagement with video-related content on social media platforms, mainly YouTube and TikTok.
O'Brien and Toms~\cite{o2008user} provide a conceptual framework to define user engagement: ``a quality of user experience characterized by attributes of challenge, positive affect, endurability, aesthetic and sensory appeal, attention, feedback, variety/novelty, interactivity, and perceived user control.''
The user engagement definition is pretty broad, covering many aspects ranging from users' immediate actions to the content (e.g., liking videos) to psychological factors (e.g., positive affect, perceived user control, etc.). 
For the purposes of our work, we mainly focus on the user engagement attributes related to attention and feedback, aiming to understand the interplay of user engagement and the recommendation algorithm on TikTok. 
}
\revision{Prior work focuses on user-generated video platforms like YouTube and TikTok to analyze user engagement.
Specifically, Yang et al.~\cite{yang2022science} study user engagement with online science videos on YouTube, focusing on how video characteristics likely affect user engagement, highlighting that users tend to view shorter videos.
Khan~\cite{khan2017social} surveys YouTube users, finding that users' motives play a significant role in user engagement in the form of liking or disliking videos on YouTube and that males are more likely to dislike YouTube videos.
Park et al.~\cite{park2016data} focus on analyzing the video view duration on YouTube (i.e., time spent on a video by users), finding that the video view duration is associated with the video's view count, the number of likes, as well as the sentiment in the comment.
Spartz et al.~\cite{spartz2017youtube} undertake an experiment on a climate change YouTube video, showing that people are more likely to engage with the video when the video has a large number of video views.
Cheng and Li~\cite{cheng2023like} focus on user engagement on news-related TikTok videos, showing that videos with negative sentiment had significantly higher user engagement.
Overall, most prior work on user engagement on user-generated video platforms like YouTube and TikTok focuses on specific types of videos (news) or reporting aggregate results based on a video's overall engagement. 
In contrast, in our work, we leverage a data donation system to obtain user behavioral traces, which allows us to perform a fine-grained analysis of user engagement without focusing on videos on a specific topic.
}\revisioncomment{R3}

\subsection{Data Donation}

Most of the prior work on TikTok detailed above uses data scraped from public content on the app, user data from researcher-created accounts, or self-reported user data. Our work diverges from previous research by studying TikTok with data donated directly by users. 
Per the EU's GDPR~\cite{gdpr}, most major digital platforms now provide their users with electronic access to the personal data they have on each user via downloadable data packages~\cite{boeschoten2020digital}. A prominent movement in the medical field~\cite{bietz2019data,strotbaum2019your}, researchers studying digital platforms are beginning to leverage the rich information in these packages by requesting that users donate them for study. Data donations offer unique insights into digital platforms~\cite{van2021promises}: for example, uncovering widely-used ad targeting mechanisms on Twitter that were largely ignored by prior work~\cite{wei2020twitter} and gaining new insights into how adolescents use Instagram~\cite{caddle2021instagram} have all been possible via user data donation.

Motivations for using user-donated data stem from the limitations of other methods. People's perceptions of their own online behavior, for instance, can be unreliable \cite{parry2021systematic,verbeij2021accuracy,burnell2021associations,ernala2020well}. Additionally, researcher-created accounts on digital platforms may lack the authenticity, diversity, and history of real user accounts. Further, scraping TikTok data yields fruitful datasets yet has capacity limitations and totally relies on public -- and typically, popular -- content available on the platform. User-donated TikTok data can provide further insight into how real TikTok users are consuming content on TikTok.

\subsection{Remarks}
\revision{To the best of our knowledge, we perform the first study that leverages data directly donated from users to study user engagement with short-format videos on TikTok and shed some light on the effectiveness of the recommendation algorithm.}\revisioncomment{R4}
Also, in contrast to previous work that focuses on understanding algorithmic recommendations using synthetic traces from automated accounts, we rely on a dataset donated directly from real TikTok users, which offers a detailed and comprehensive view of algorithmic recommendations made to users, as well as their actions (e.g., liking videos, time spent on each video, etc.) on TikTok.
We argue that using data from real users is paramount, as synthetic traces from automated accounts lack the authenticity and diversity of real user accounts.

\section{Data Donation System}\label{sec:platform}

We implement a data donation system, \emph{Social Media Donator (SMD)}, where users can get information on how they can request their data from the TikTok mobile application.
In this section, we provide details on how users can request their TikTok data and what is included in the data.
After users download their data, they can use SMD to anonymize and customize their data \emph{before} transferring it to our backend.

\subsection{Requesting Data from TikTok}\label{sec:requesting-data}
TikTok enables its users to request a comprehensive dataset of their activity on the platform and other personal information the platform has on them. The request can be made through the TikTok mobile application's settings menu, and the data will be provided in either JSON or a human-readable format based on the user's preference. The process takes a few days for the data to be ready for download. Here, we outline the various fields of information included in a user's TikTok data download.

\begin{itemize}
    \item \textbf{Video Viewing History:} A record of the videos the user watched and the time they began watching them.
    \item \textbf{Like History:} A record of the videos the user liked and the time they liked them.
    \item \textbf{Search History:} A record of the search queries the user made on TikTok, along with the time they made each query.
    \item \textbf{Share History:} A record of the videos the user shared, the time they shared them, and the method used to share (e.g., WhatsApp, Facebook Messenger, etc.).
    \item \textbf{Login Information:}  A record of each time the user logged into the TikTok app, including the time, IP address, device and network information, and carrier.
    \item \textbf{App Settings:}  Information about the user's preferences and settings in the TikTok app, such as their interests and privacy settings.
    \item \textbf{Comments:} A record of the comments the user made, along with the time they made them. Note that the file does not include the specific videos on which the comments were posted.
    \item \textbf{Favorites:} A record of the videos, effects, hashtags, sounds, and videos the user marked as favorites on TikTok, along with the time they marked each item as a favorite.
    \item \textbf{Following/Followers:} A record of the accounts the user follows/who follow the user and the time of the follow action.    
    \item \textbf{Ads Information:} Information about the advertisers that targeted the user.
    \item \textbf{Profile Information:} Information the user provided about themselves in their TikTok profile, such as their bio, email address, phone number, username, and profile photo.
    \item \textbf{Direct Messages:} A record of private messages the user exchanged with other TikTok users.
    \item \textbf{Video Uploads:}  A record of the videos the user uploaded to TikTok.
    \item \textbf{Purchase History:} Information about purchases the user made within the TikTok platform.
    \item \textbf{Account Status:} Information about the status of the TikTok app on the user's device, such as the app version and screen resolution.
\end{itemize}

\subsection{Data Anonymization.} The data collected from TikTok includes personal information and identifiers for each user, such as phone numbers and email addresses. 
Due to this, it is essential to ensure proper anonymization of the data before transferring it to our infrastructure. 
To ensure the data is properly anonymized before it is transferred to our infrastructure, our SMD system removes certain information by default. This includes the user's profile information, direct messages, information about videos uploaded, IP addresses and device information, purchase information, and account status. The anonymization process is done on the client side, and we emphasize that we only transfer the anonymized dataset to our backend. Additionally, we provide a Python script that allows users to anonymize and customize their dataset offline without using SMD. This script is identical to the one available in SMD and is intended for use by participants who are concerned about privacy.

\subsection{Data Customization.} TikTok users may have different comfort levels in sharing certain data fields. For example, a user who frequently posts comments with personal information may not feel comfortable sharing their comment-related data. To address this, we have implemented a customization feature in SMD, which allows users to choose which fields of their data they are comfortable donating. The only mandatory field is the video viewing history, which includes only the URLs of the videos watched and the timestamps. Additionally, there are some fields that users are not able to donate, as they contain personal information and identifiers (as outlined in the data anonymization procedure).
For the remaining fields that a user can choose to donate, we provide clear explanations of what data is included, with specific examples and a description of how we plan to use each field in our analysis. Additionally, for data fields that may contain sensitive information, we have added warning labels to alert users that the field may potentially include private information. For example, for the search history, we added a warning label that says, "This information may be sensitive if you did uncommon searches for things related to your real identity, e.g., searching for videos of a family member's small sports team." Similar warnings were added for the followers, following, and comments data since these fields may reveal the user's identity through their follower network and comments made on public videos.

SMD calculates the compensation for the user based on their selections of which data fields they opt-in to donate. The mandatory video viewing history is compensated with \$5, while all the optional fields such as Like History, Search History, Share History, Login Information, App Settings, Comments, Favorites, Following, Followers, and Ads Information are compensated with \$1 each, except for comments. For comments, users can either donate their comment timestamps and content for \$2 or only the timestamps for \$1. The total compensation for each user ranges from \$5 to \$16, depending on their selections.

\subsection{Data Donation \& Survey.} Users can donate and transfer their anonymized and customized data to our infrastructure with a single click on the SMD interface. After the data donation, we present all users with an optional survey that includes general demographic questions and questions about their usage of the TikTok platform and their perceptions of the TikTok algorithm's recommendations. This survey helps us to gain extra context on the users, such as their age, gender, and location. It is important to note that all questions in the survey are optional, and users can choose not to answer by selecting the "Prefer not to say" option. All users who choose to fill out the survey will receive an additional compensation of \$4 regardless of the questions they choose not to answer.

\section{Data Collection}\label{sec:data_collection}

In this section, we present our approach to collecting data from TikTok users. We describe our recruitment process, metadata collection for TikTok videos, our efforts to assess the quality of donated data, and our user viewing duration inferences. We also discuss our ethical considerations when collecting and analyzing data.

\subsection{User Recruitment}\label{sec:recruitment}

We recruited participants by 1)~sharing the study on Twitter and
2)~running Facebook Ads targeting people who are aged over 18 years old and who live in the U.S. whom Facebook had tagged with the "TikTok" interest category. For Twitter, we shared the study via a single tweet that was retweeted by all the authors' Twitter accounts. The tweet received 64.5K impressions and was shared in January 2022. For Facebook, we ran ads between January 21, 2022, and February 13, 2022, with an average budget of \$8.5 per day. 
By leveraging these two ways, we recruited 347 participants, whom we compensated with an overall amount of \$6.9K in the form of Amazon gift cards sent via email.
Note that we do not have a way to distinguish recruited participants that participate in the study because of the Facebook ads, Twitter, or possibly hearing about the study advertisement elsewhere. %
\revision{Also, we note that all recruited participants were active users on TikTok by the time of our data collection. Hence we are unable to study the dynamics between retained and lost users on TikTok (i.e., users that stopped using the platform).}\revisioncomment{R4}

\begin{figure}[t]
\centering
\includegraphics[width=\columnwidth]{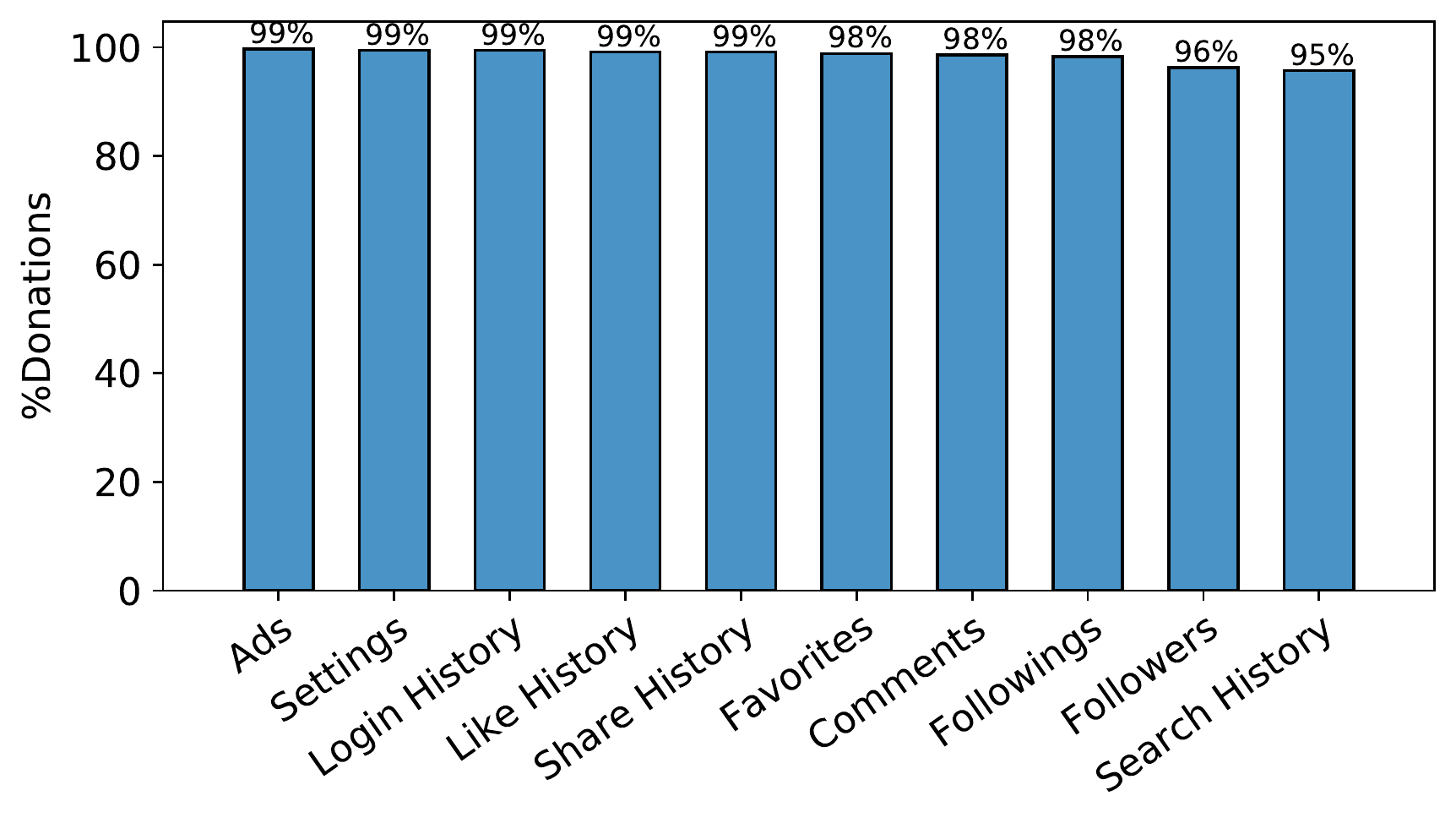}
\caption{Percentage of donations that opt-in to donate to each field included in the TikTok data.}
\label{fig:donation_options}
\end{figure}

Fig.~\ref{fig:donation_options} shows the percentage of participants that opted-in to donate each potential field that exists in their TikTok data. 
As we can observe, most of our participants chose to donate almost all the fields, as all fields appear in at least 95\% of all the donations.
Participants were less willing to share their Search History (95\%) and Followers (96\%), Following (98\%), and Comments (98\%).
This is potentially due to the warning labels associated with these fields in our SMD interface, explaining that some information included in these fields might be sensitive (see Section~\ref{sec:platform}).
This result suggests that most participants perceive the trade-off between the compensation and the donation of additional fields as worthwhile.

\descr{Participants' Demographics.} 
96\% of all participants filled out our survey, hence we obtained demographic information about them.
Regarding the location of our participants, we found that slightly more than half were from Africa (52\%). The rest came from North/Central America (32\%), South America (6.6\%), and Europe (3\%). Notably, a significant portion of our participants were from Africa, even though our recruitment efforts primarily targeted Facebook users in the U.S. We speculate that the financial incentives offered may have been particularly attractive to participants from Africa, suggesting that individuals from underdeveloped countries are willing to contribute their data for research in exchange for monetary compensation.
In terms of the age of our participants, we found that the vast majority were 34 or younger (91\%), with almost half (48.6\%) between 25-34 and slightly less than half (43\%) aged between 18-24.
Our participants set is somewhat gender-balanced, with 55\% of the participants being men, while 43.2\% of our participants are women and 1.2\% are non-binary or self-described their gender. 
In terms of education, the majority of our participants have attained a higher level of education, with over half (54\%) holding a bachelor's degree or higher (such as a Master's or PhD), while a third (32\%) have completed post-secondary education or have an associate's degree or equivalent.

\begin{table}[t]
\centering
\small
\begin{tabular}{@{}lrr@{}}
\toprule
\textbf{}                      & \multicolumn{1}{l}{\textbf{\#Actions}} & \multicolumn{1}{l}{\textbf{\#Participants}} \\ \midrule
\textbf{Video Viewing History} & 9,212,100  & 347                                                                      \\
\textbf{Like History}           & 1,120,716  & 328                                                                    \\
\textbf{Search History}        & 13,282  & 332                                                                        \\
\textbf{Share History}         & 24,944   & 253                                                                       \\
\textbf{Comments}               & 52,436 & 227                                                                        \\
\textbf{Following}            & 84,654 & 333                                                                          \\
\textbf{Followers}             & 43,642   & 295                                                                       \\ \bottomrule
\end{tabular}%
\caption{Dataset statistics.}
\label{tab:dataset-overview}
\end{table}

\descr{Dataset.} Table~\ref{tab:dataset-overview} provides statistics about our dataset, which includes 9.2M video views made between July 26, 2020, and February 21, 2022.
Our dataset also includes other actions made on the platforms, particularly 1.1M like actions from 328 participants, 13K search actions from 332 participants, 24.9K shares from 253 participants, 52.4K comments from 227 participants, 84.6K following actions from 333 users, and 43K follower actions from 295 participants. 
Additionally, upon analyzing the temporal distribution of these actions in our dataset, we discovered a 2-month gap in the data for likes. Specifically, there is no information on likes for any of the 347 recruited participants, which we suspect is due to technical difficulties with the data logging infrastructure within TikTok. Furthermore, we found that data on video sharing is not present until July 28, 2021.

\subsection{Video Metadata Collection}

Each participant's data contains information on their activity, such as the videos they viewed, liked, shared, etc. It's worth mentioning that for each video, the data only includes the video's URL. To gain additional context and information about the TikTok videos included in our donations, we utilized an unofficial Python API wrapper~\cite{unofficial_wrapper}, which uses Selenium to extract the metadata of each video in JSON format by scraping the TikTok page. We found 4,938,805 unique TikTok videos viewed by our participants among the 347 donations. We attempted to gather metadata for all the videos, successfully obtaining it for 4,122,038 videos (83.4\% of all videos). The remaining videos were either deleted by the uploader or by TikTok, or the account that posted them had set their account to private. As we collect participants' data donations that include their entire activity, we also collected videos from 2020 that are more likely to be deleted when compared to newer videos.
We collected video metadata between January 17, 2022, and March 12, 2022.
For each video, we collected the following metadata:
1) Date and time of the video creation;
2) A description and a title for each video as defined by the uploader;
3) Uploader-specific information like the uploader's username and unique identifier;
4) Video metadata like the duration, format, etc.; and
5) Video engagement statistics on the entire TikTok platform, such as the number of shares, comments, and number that the video was viewed (at the time of our data collection).

\subsection{Assessing the ``Quality'' of Donations} SMD requires participants to provide an email for the purposes of sending the compensation.
We also use the MD5 hash of each email as the unique identifier for each donation. Note that we do not explicitly link the email address with the donation for anonymity reasons.
As expected with studies that offer a monetary incentive, users may try to ``trick'' the system to earn money easily.
Indeed, during our recruitment, we noticed that some TikTok users were trying to earn more money by donating duplicate or near-duplicate data under different email addresses.
To detect such malicious donations, SMD calculates all the pairwise Jaccard similarities between the video URLs and timestamps that are included in the video viewing history.
Then, SMD flags the donations with a Jaccard similarity of over 0.2 for either the video URLs or the timestamps, and then we manually check the donations to verify that they are indeed duplicate or near-duplicate donations. This process is done before sending the compensation.
Overall, we received 31 duplicate or near-duplicate donations (all of them having a Jaccard similarity of 0.9 or more) that would have cost us \$571. 
For these donations, we did not pay users, and we informed them via email that they would not receive compensation for all their subsequent duplicate donations (as they had already received compensation). 

Another concern is that malicious users can generate fake video URLs and timestamps and try to donate fake data for compensation. 
To verify if this is happening in our recruitment, we check for what percentage of each donation's videos we obtained video metadata.
Here, we hypothesize that if a malicious user generates fake URLs, then we will not be able to get any metadata for the videos included in the donation (since the URLs will not exist).
We did not find any evidence of users submitting fabricated donations since, for all the donations, we were able to obtain metadata for a large percentage of the videos --- at least 70\% for every donation, with a median of 90\%. As discussed above, the rest of the videos were inaccessible because they were either deleted or the account became private.

\begin{figure}[t]
\includegraphics[width=\columnwidth]{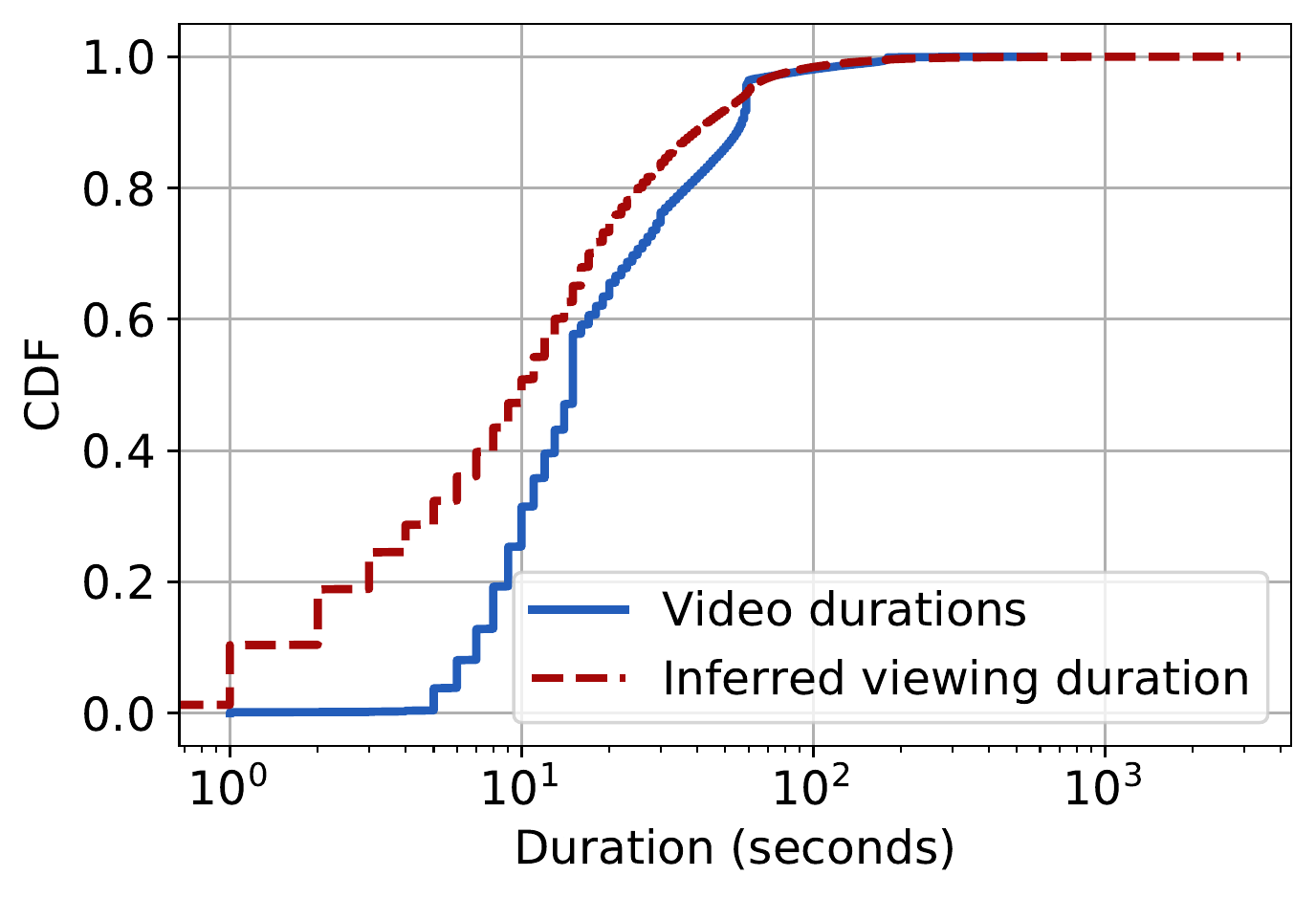}
\caption{CDF of video durations/inferred viewing duration.}
\label{fig:cdf_video_durations}
\end{figure}

\subsection{Inferring Viewing Duration}\label{sec:inferring}
Our dataset includes the video URLs and when each user started watching each video.
For the purposes of our study, we want to assess whether users watched videos until the end, hence we aim to infer the duration of each video view based on the timestamps.
We use the methodology by Halfaker et al.~\cite{halfaker2015user} that performs user session identification based on the inter-arrival times of the user actions.
Specifically, we first calculate the time difference between consecutive timestamps of video views.
This allows us to approximate the video view duration, except when the user takes a break (the duration will be unreasonably high).
The method by Halfaker et al.~\cite{halfaker2015user}, then uses a two-component Gaussian mixture model that performs clustering on the inferred video view durations.
The main idea is that the first cluster will include all the durations corresponding to actual video views, while the second cluster will include all inferred durations corresponding to user breaks.
The clustering allows us to identify a threshold by assigning the video view durations to the clusters.
Based on this method and the dataset, we use a threshold of 105 seconds.
We believe that this threshold is reasonable as the Cumulative Distribution Function (CDF) of all the inferred viewing durations and all video durations (obtained from the video metadata) in Fig.~\ref{fig:cdf_video_durations} shows that 98.5\% of all the inferred viewing durations are 105 seconds or less.

\revision{Having inferred the viewing duration for each video view in our dataset, we then calculate an \emph{Attention} metric, which is simply the percentage of viewing duration based on each video's duration (we divide the inferred duration with the video's duration, and we multiply by 100). The Attention metric allows us to assess if a user watched a video until the end (i.e., if the metric is $\ge 100$\%), whether a user watched the video multiple times (i.e., if the metric is $\ge 200$\%), or whether a user continued to the next video before the video's completion (i.e., if the metric is $<100$\%).}\revisioncomment{R1}

\subsection{Ethics}\label{sec:ethics}

Before recruiting participants and gathering data from real TikTok users, we received approvals from two Ethical Review Board committees, one from Saarland University and one from the University of Washington. We submitted a comprehensive document to the committees outlining the various data fields included in the TikTok data, our methods for anonymization and customization, the consent form that would be presented to users, and our recruitment advertisements.
For each participant, we obtained explicit consent by providing them with a consent form that they can download through our SMD donation system and by ensuring that they understand and agree to the terms outlined in the form. Additionally, as outlined in Section~\ref{sec:platform}, we explained to participants how we planned to use the data they donated and the potential privacy implications of providing certain data fields (such as comments, search history, and follower/following network). As previously mentioned, all users who donated their data received compensation in the form of Amazon gift cards valued between \$5-\$20, depending on the data fields they chose to donate and whether they completed the optional survey.

We will permanently delete all user data within 36 months after the completion of our project, and the data will not be shared with any third parties. Additionally, we want to stress that participants have control over which data fields they choose to donate. We strictly abide by standard ethical guidelines~\cite{rivers2014ethical} in our analysis, such as reporting our results in aggregate and not attempting to track users across platforms. Additionally, our metadata collection is limited to publicly available videos on the TikTok platform during the data collection period, thus, we do not collect any information about private or deleted videos.

\section{Results}\label{sec:results}

This section presents our analysis and results that aim to assess the effectiveness of TikTok's recommendation algorithm through the lens of user engagement.

\subsection{\revision{Characterizing our participants' viewing activity on Tiktok and how it changes over time}\revisioncomment{R1}}\label{sec:high-level}
We start with a high-level characterization of the dataset, focusing on the participants' activity on TikTok and how it changes over time. We look into how much time the participants spent on TikTok, how many videos they watch daily, how many videos are from following accounts, and how these metrics change over time.

\descr{Time Spent.} \revision{To calculate the time spent, we sum all the inferred viewing durations (in seconds) except the last video of each session, as we can not reliably determine how much time a user spent on the last video of each session (see Section~\ref{sec:inferring}).}\revisioncomment{R1} We find a median value of 1,622 seconds per day (27 minutes), with 25\% of the participants (\emph{Q1}) spending on average less than 834 seconds per day (13.9 minutes) on TikTok, while the 25\% most active (\emph{Q3}) participants spent 2,891 seconds (48 minutes) or more ($\sigma$: 1,864 seconds). 
This consumption is larger compared to other video platforms like YouTube, where people spend on average 19 minutes~\cite{youtube_use_stats}.

\descr{Volume of Videos.}
We find that, on average, the participants watch a substantial number of videos per day, with a \emph{median} of 89.9 videos per day (\emph{Q1: 40.7, Q3: 170.3 }, $\sigma$: 128.9).
This is somewhat expected since TikTok videos are usually short in length, hence users can watch a large number of videos without spending much time (compared to other video platforms like YouTube).

\begin{figure}[t]
\subfigure[Volume]{\includegraphics[width=\columnwidth]{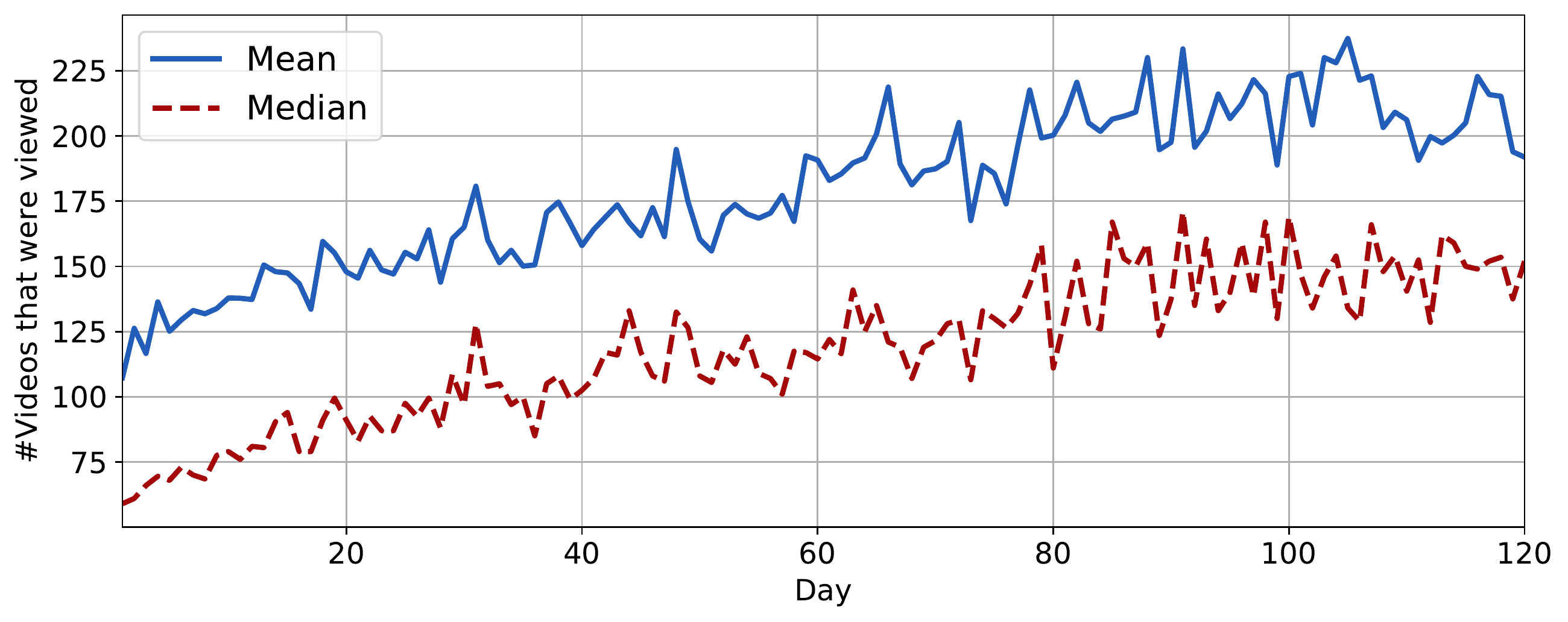}\label{fig:tenure-evolution-views}}
\subfigure[Time]{\includegraphics[width=\columnwidth]{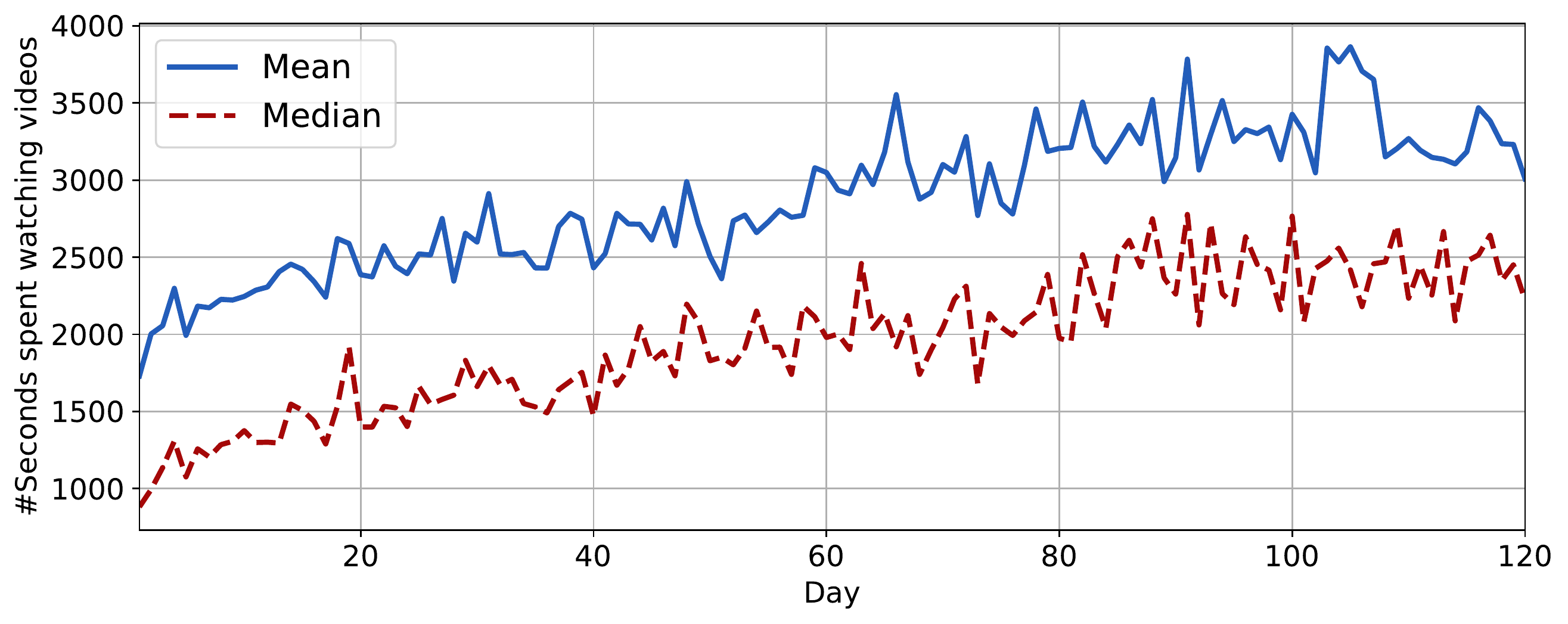}\label{fig:tenure-evolution-time}}
\caption{Changes in the volume of watched videos and time spent on TikTok over time. We observe that over time, there is an increase in both the volume of videos watched per day and the daily time spent on TikTok.}
\label{fig:tenure-evolution-consumptive}
\end{figure}

\begin{figure}[t]
\includegraphics[width=\columnwidth]{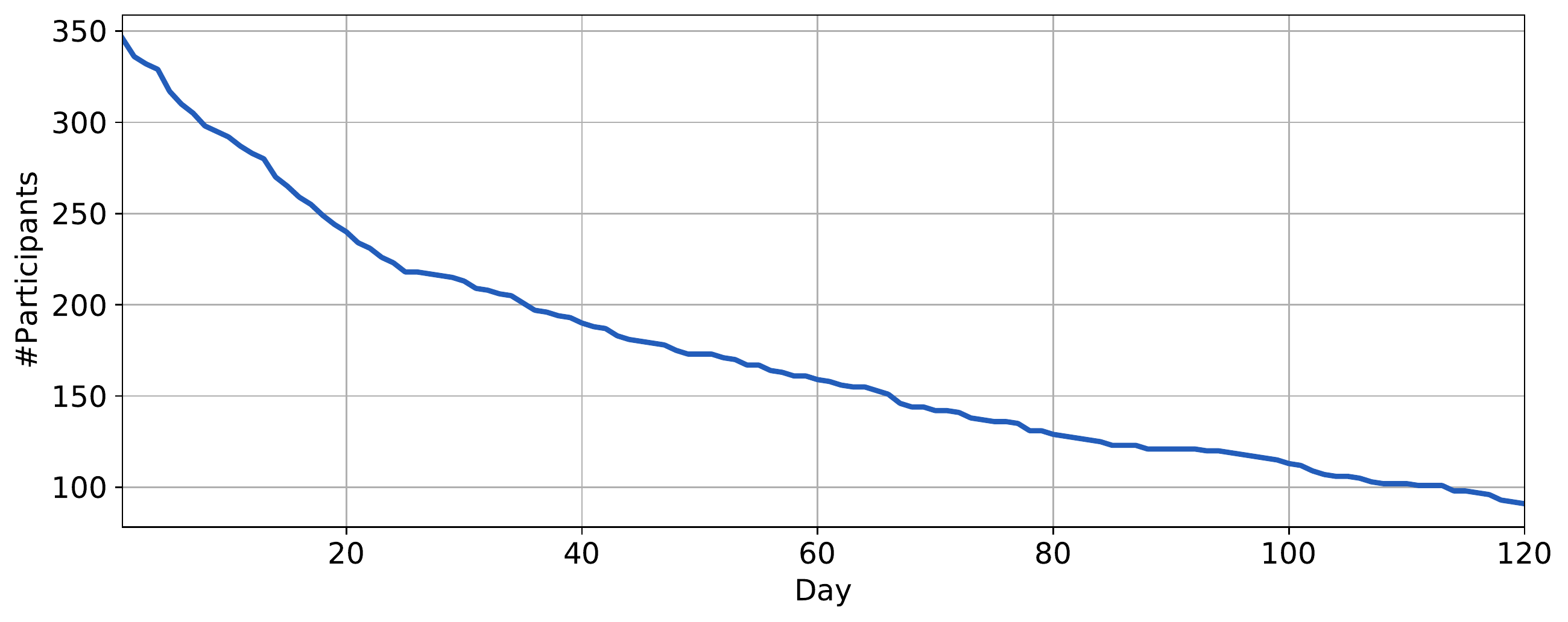}
\caption{Number of participants that are considered for each day in our temporal analysis. The first day considers all participants while the 120th day considers 26\% of them.}
\label{fig:samples-tenure}
\end{figure}

\descr{The TikTok participants view more videos and spend more time on the platform over time.} Here, we aim to longitudinally analyze how the volume of videos and time spent on TikTok changes over time for the participants.
We do this analysis to understand how compelling the TikTok platform is over time (according to the participants' data), which gives us some insights into whether the recommendation algorithm achieves its goal of keeping people watching videos.
We undertake a temporal analysis where we report the aggregation of the time spent and volume of videos over time by normalizing the time across participants. First, we split the viewing histories of all participants into equal periods (i.e., per day). Then, we align the viewing histories across participants by considering the index of the day (rather than the absolute day). We do this alignment mainly because the participants started using TikTok at different time periods.
We calculate each metric and report the aggregate results by reporting the mean and median values of these metrics across all participants for each relative day.
\revision{Note that our analysis, for each participant, focuses only on days when a participant was active on TikTok. For instance, if a user started watching videos on the first day, then had two days of break (i.e., not using the platform), and then returned on the fourth day, then we assume that the fourth day is the participant's second active day, which is what is represented in our analysis.\footnote{\revision{While investigating these gaps is important for understanding user retention on the platform, we leave this analysis as part of our future work.}\revisioncomment{R1}} This is a crucial preprocessing step since keeping days when participants are inactive will lead to underestimating the aggregate activity (e.g., the median number of watched videos is zero simply because most of the participants were not active during the same normalized days).}\revisioncomment{R1}

Fig.~\ref{fig:tenure-evolution-consumptive} provides an overview of how the volume of videos and time spent on TikTok changed over time.
We limit the analysis to the first $120$ days because, over time, the number of participants that are analyzed decreases (see Fig.~\ref{fig:samples-tenure}, this is because the participants have varying tenure duration on TikTok), and we wish to avoid making conclusions based on only a small number of participants.
From Fig.~\ref{fig:tenure-evolution-consumptive}, we can observe that the mean/median volume and time spent on the platform increase over participants' tenure. 
For instance, the average number of viewed videos starts at 107.31 videos per day, while after 80 days, it is consistently over 200 videos per day, with a peak of 233 videos per day.
The same applies to the time the participants spent on the platform; on the first day, the average number of seconds is 1,730 (nearly 29 minutes), while after 120 days, it is over 3,000 seconds (50 minutes) per day. 
To assess the effect of the decreasing number of participants over time, we repeat the same analysis with only the participants that are active for all 120 days (26\% of all participants). Our analysis shows that the results presented above are not due to the decreasing number of participants over time but are based on the participants' behavior changes.

\begin{figure}[t]
\subfigure[Number of followings]{\includegraphics[width=\columnwidth]{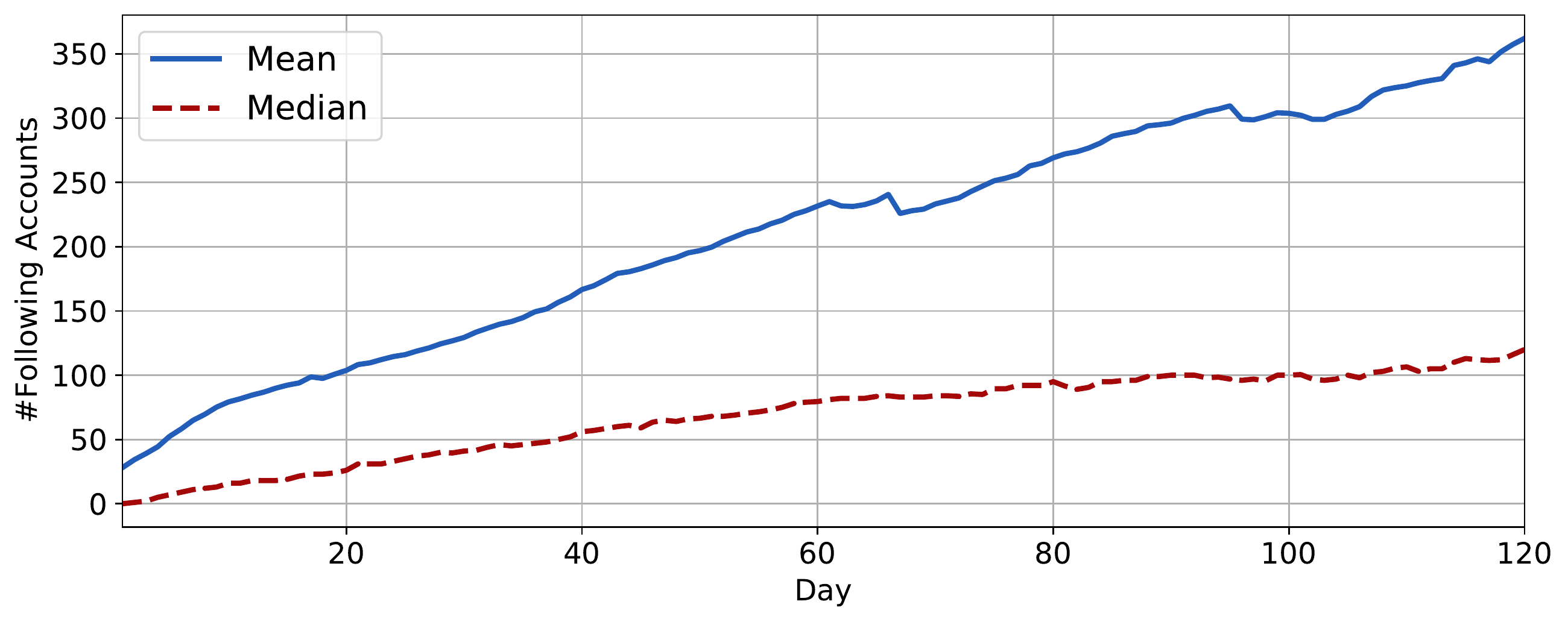}\label{fig:following_accounts_day}}
\subfigure[Views for videos uploaded by followings]{\includegraphics[width=\columnwidth]{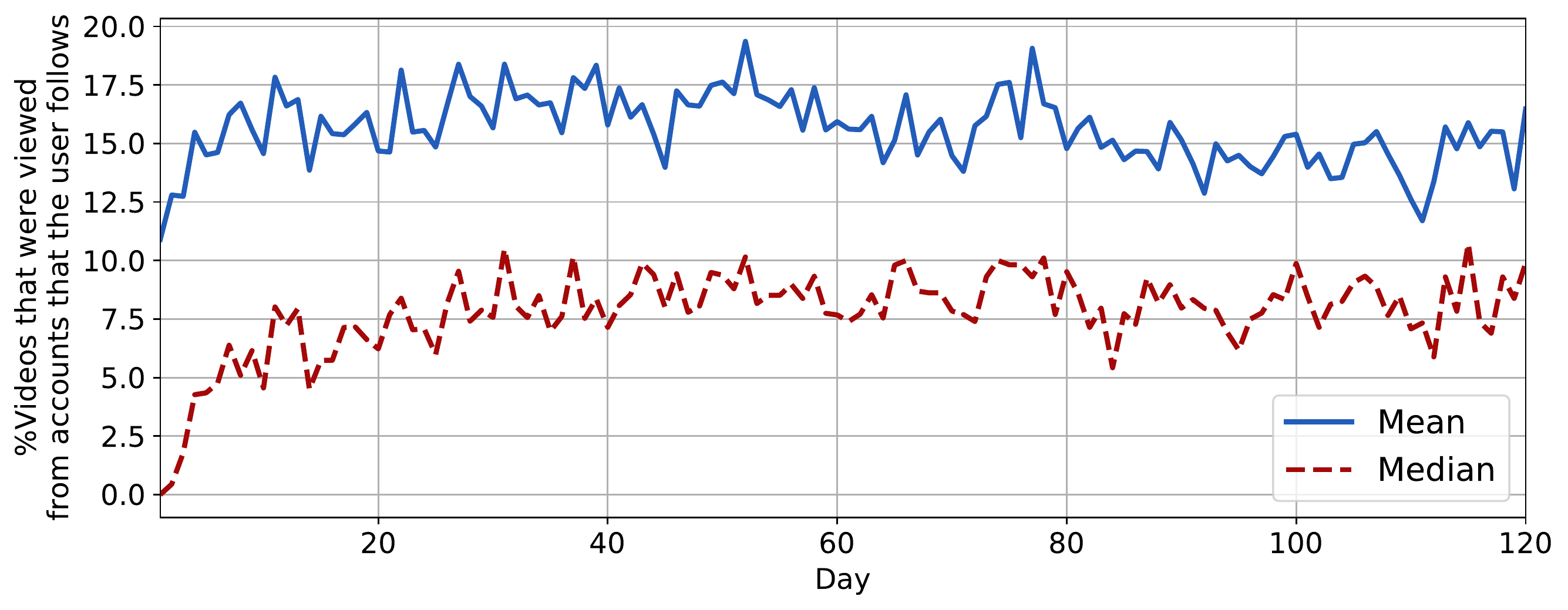}\label{fig:following_views_day}}
\caption{Number of following accounts and number of videos that are viewed from following accounts over time. The aggregate number of following accounts increases over the users' tenure, while the percentage of videos that are watched from following accounts remains stable over the users' tenure.}
\label{fig:tenure-evolution-followings}
\end{figure}

\descr{Following.} \revision{During the early days of typical social media platforms like Facebook and Twitter, most of the content consumed by the users originated from users they follow.}\footnote{\revision{Note that in recent days, platforms like Facebook and Twitter are becoming more TikTok-alike, as a large portion of content that appears to user feeds originates from outside their social networks.}\revisioncomment{R4}}\revisioncomment{R4}
TikTok has a different approach to delivering content since the ``For You Page,'' consists of a lot of videos coming from accounts that a user does not follow.
We investigate this phenomenon through the lens of the dataset, mainly aiming to assess how prevalent is the consumption of content coming from the following accounts and how it changes over time.
To do this, for each participant, we extract the accounts that the user follows (along with the associated timestamp of the following action) and then label all videos from those accounts as videos from followings, only if the video view timestamp is after the following action.
Overall, we find that only 10.3\% of the video views in the dataset are actually for videos originating from accounts that the participants followed in advance, emphasizing that the majority of content consumed on TikTok is originating from accounts that the participants did not explicitly follow.
Also, we assess changes over time by visualizing the number of accounts that the participants follow over time, as well as the percentage of all video views originating from following accounts, per participant (see Fig.~\ref{fig:tenure-evolution-followings}).
As expected, over time, the participants follow more and more accounts, hence the number of following accounts grows over time (see Fig.~\ref{fig:following_accounts_day}, on average 40 accounts on day 1, while after 120 days, the average is 350). 
However, when we look into the number of video views from the following accounts (see Fig.~\ref{fig:following_views_day}), we observe that the number of videos from the following accounts that the participants consume is stable over time (a median of 10\% of the videos that a user consumes are from the following).
This likely indicates that the TikTok algorithm or the platform features are designed in such a way so that the users receive only a limited number of videos from following accounts, irrespective of the number of accounts a user follows.

\descr{Take-Aways.} The main take-away points from our high-level characterization are:
\begin{itemize}
\item We find that the participants watch a larger number of videos over time, resulting in spending more time watching videos on TikTok (e.g., on average, participants' time spent increased $2\times$ after 80 days).
\item Despite the fact that the participants follow more accounts over time, the percentage of videos that they watch and are from following accounts is stable over time, with a median of approximately 10\% of all video views.
\end{itemize}

\subsection{\revision{Analyzing our participants' engagement activity and its evolution over time} \revisioncomment{R1}}

This section analyzes the participants' engagement behavior. We focus on: 1) users' attention (i.e., watching the video until the end); and 2)~users' interaction through video liking.
\revision{While we acknowledge that comments and shares are crucial engagement signals, our analysis focuses only on likes for various reasons. First, for comments, our dataset does not include the video in which a comment was made; hence, using comments data is not trivial and will require inferences based on timestamps. The accuracy of such inferences will be hard to measure; hence, we refrain from using comments in our analysis. Second, given that shares comprise only 2\% of all engagement actions, we elected to focus only on likes as the shares will not yield any statistically significant differences in our results.}\revisioncomment{R4}

\descr{What percentage of each video did the participants watched on TikTok?} We take into account the video duration and how much time participants spend watching each video (relative to its duration) to compute an \textit{Attention} metric. Measuring attention allows us to understand how many videos are watched to the end or skipped before the end. We compute the \textit{Attention} metric by calculating the inferred viewing duration divided by the video duration (obtained from the video metadata) and multiplied by 100.
If this percentage is 100\%, it means that the participant watched the video until the end. If it is less than 100\% it implies the participant skipped the video by swiping up before the end of the video, and if the percentage is greater than 100\% it means that the participant watched the video (or portions of it) multiple times (in these cases we set the Attention of that video to 100\% since the participant watched the entire video).
Note that our \textit{Attention} analyses are based only on video views for which we have video metadata (83.4\% of all videos).

\begin{figure}[t]
\includegraphics[width=\columnwidth]{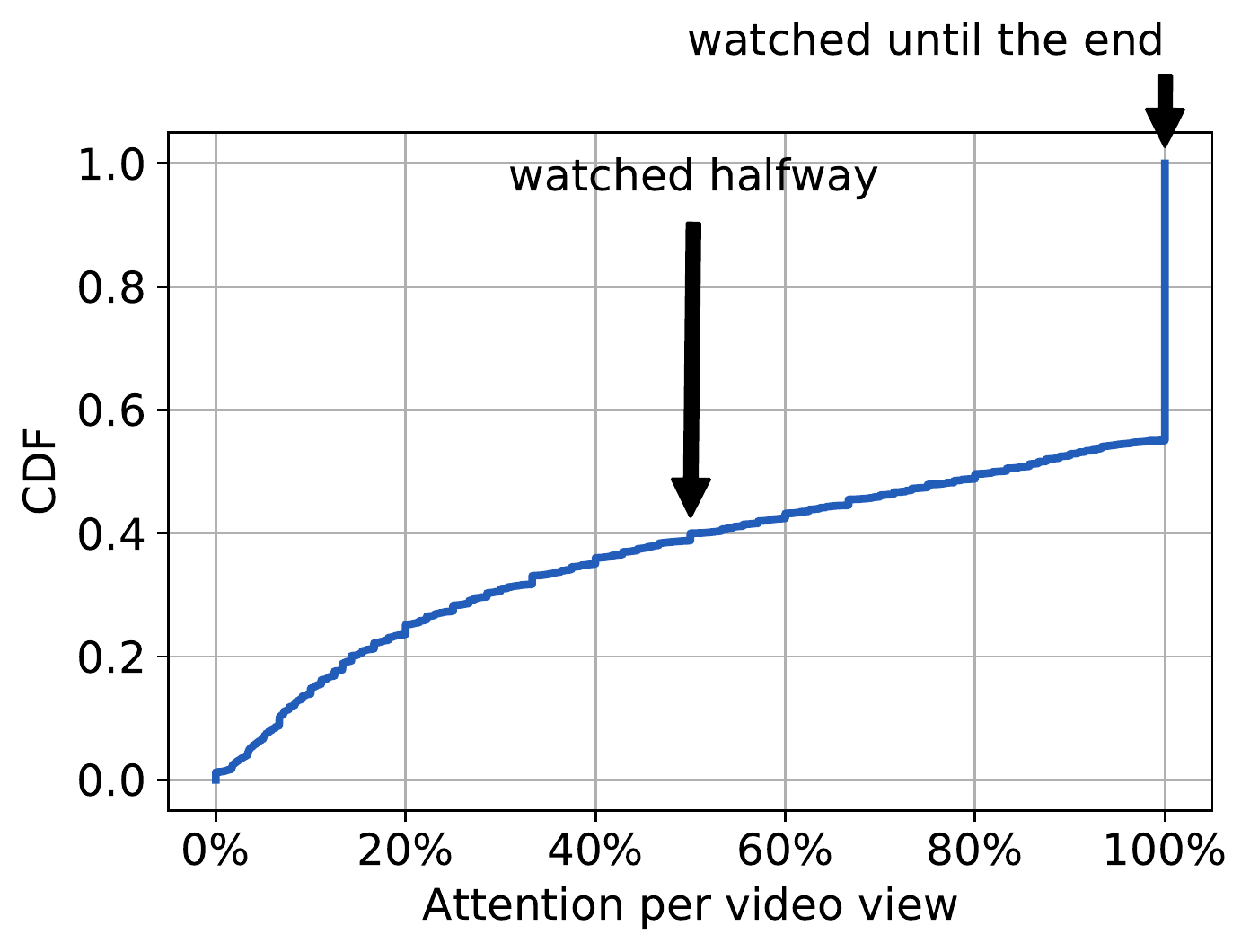}
\caption{CDF of the participants' attention per video view. For 45\% of the video views, the participants watched the video until the end.} 
\label{fig:cdf_viewing_percentage_all}
\end{figure}

\begin{figure}[t]
\includegraphics[width=\columnwidth]{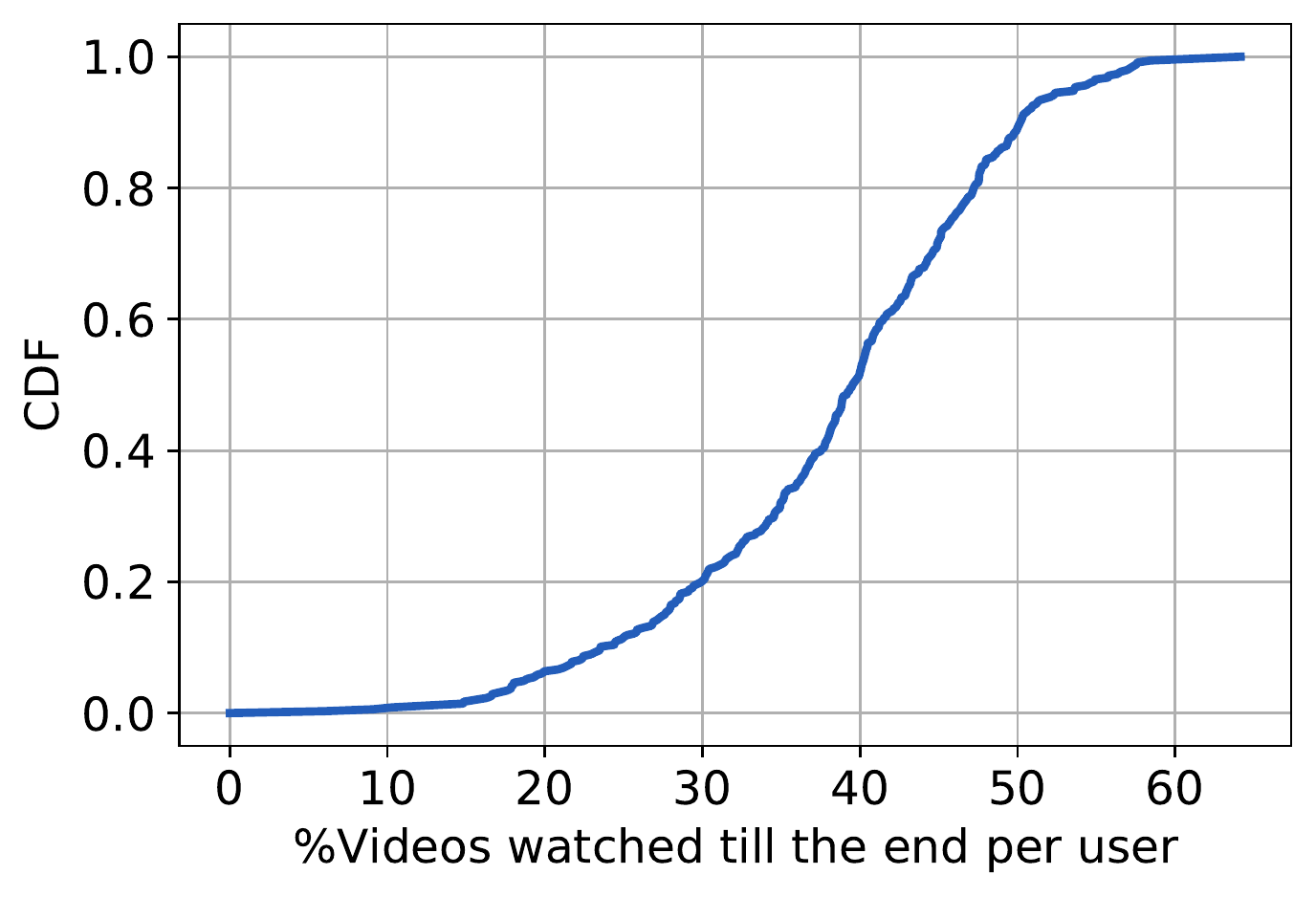}
\caption{CDF of the percentage of videos watched until the end, per participant. The majority of the TikTok users in the dataset watch until the end between 20\% and 60\% of all videos they watched.} %
\label{fig:cdf_percentage_until_end_per_user}
\end{figure}

Fig.~\ref{fig:cdf_viewing_percentage_all} shows the CDF of the participants' attention for each video view.
We find that the median attention is 82\%, which indicates that the participants watch most of the video's duration before going to the next video.
Also, we find that for 45\% of the video views, the participants watched the videos until the end, while for the rest 55\% of the video views, the participants proceeded to the next video before its completion.
In particular, 24\% of the video views are skipped quickly (i.e., before watching 20\% of the entire video duration), and 40\% of the video views are skipped before watching 50\% of the video.
By aggregating the results on a participant level (see Fig.~\ref{fig:cdf_percentage_until_end_per_user}), we find that no participant watched until the end more than 65\% of the videos, while 70\% of the participants watched until end 30\%-50\% of all the videos in their watching history.
The fact that a large number of videos are skipped before their end either indicates that recommending short-format videos to users is a hard task or that the TikTok algorithm recommends videos that are likely to be skipped to maximize users' long-term excitement and retention on the platform.

\descr{The TikTok participants have stable Attention over time and they tend to watch more videos until the end for videos from non-following accounts.} 
Here, we aim to understand how the participants' attention changes over time, as this analysis will indicate how well the TikTok algorithm recommends content that participants watch to the end.
For instance, if the algorithm effectively infers participant interests, we expect that the percentage of videos that are watched to the end will increase over time.
\revision{We assume that there is no ceiling effect (i.e., users having a limited time to view videos), as based on our analysis, we find that on aggregate, users' time on the platform increases over time, and we do not observe any ceiling effect (see Fig.~\ref{fig:tenure-evolution-time})}\revisioncomment{R4,R2}.
Also, this analysis provides us with a sense of how compelling TikTok is for the participants, as we can assess how attention changes over time. 
We perform a similar analysis to the one presented in Section~\ref{sec:high-level}, particularly looking into how the Attention (percentage of videos watched until the end) signal changes over time for the participants, separating videos originating from followings and non-followings. 
Fig.~\ref{fig:tenure-evolution-attention} shows how the Attention signal changes over the first 120 days of the participants' tenure.

We make several observations. 
First, we find that the participants' attention (i.e., the percentage of videos watched till the end) over time is somewhat stable, which is surprising since we expected it to increase over time.
This is because previous anecdotal~\cite{tiktok_algorithm_good} evidence suggests that the TikTok algorithm performs well in inferring users' interests and recommending engaging content.
Second, we observe that videos coming from non-followings attract more attention from the participants compared to videos coming from followings (between 44\% and 46\% of the videos are watched until the end for videos from non-followings, while the same percentage is between 38\% and 42\% for videos from following).
This finding might also explain why TikTok limits the videos from following accounts (see Fig.~\ref{fig:following_views_day}). 
In other words, TikTok participants are less likely to watch until the end videos from followings, and since the goal of the algorithm is to recommend videos that people will eventually watch till the end, TikTok is limiting the prevalence of videos from following accounts in its recommendations.
\begin{figure}[t]
\includegraphics[width=\columnwidth]{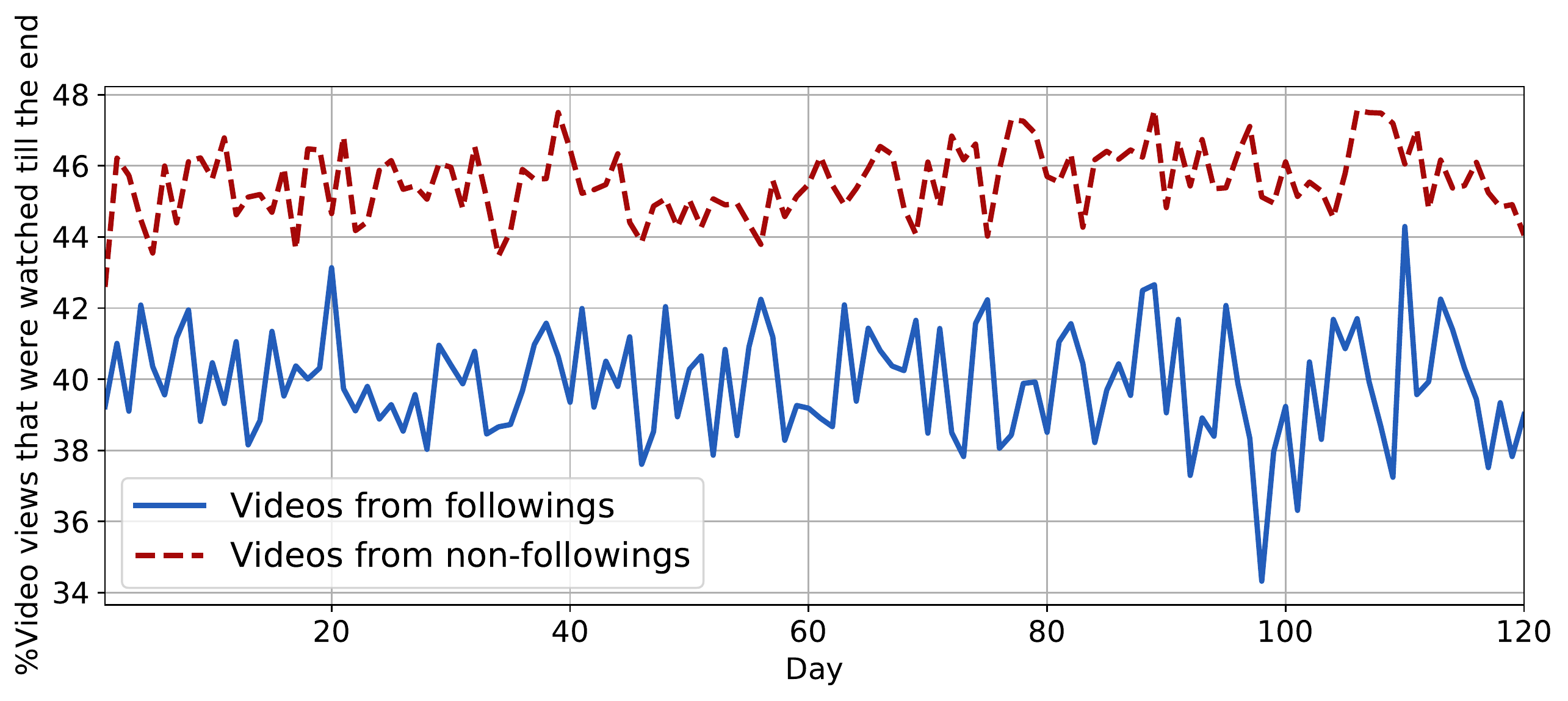}\label{fig:tenure-evolution-attention}
\caption{Mean attention (percentage of videos watched until the end) over time. We find that the attention metric does not substantially change over the users' tenure on TikTok.}
\label{fig:tenure-evolution-attention}
\end{figure}

To assess the robustness of our results, we make a sensitivity analysis concerning the number of participants included in the analysis and the threshold used for watching a video until the end (we used 100\%).
With regard to the effect of the varying participant numbers over time, we repeat the analysis considering only the participants that were active for all 120 days.
We find a similar pattern as the attention does not change substantially over time.
Also, we repeat the analysis, with a varying threshold for watching a video until the end (between 50\% and 90\%), finding that our results still hold even when considering a less strict threshold for inferring when a video is watched until the end.

\begin{figure}[t]
\includegraphics[width=\columnwidth]{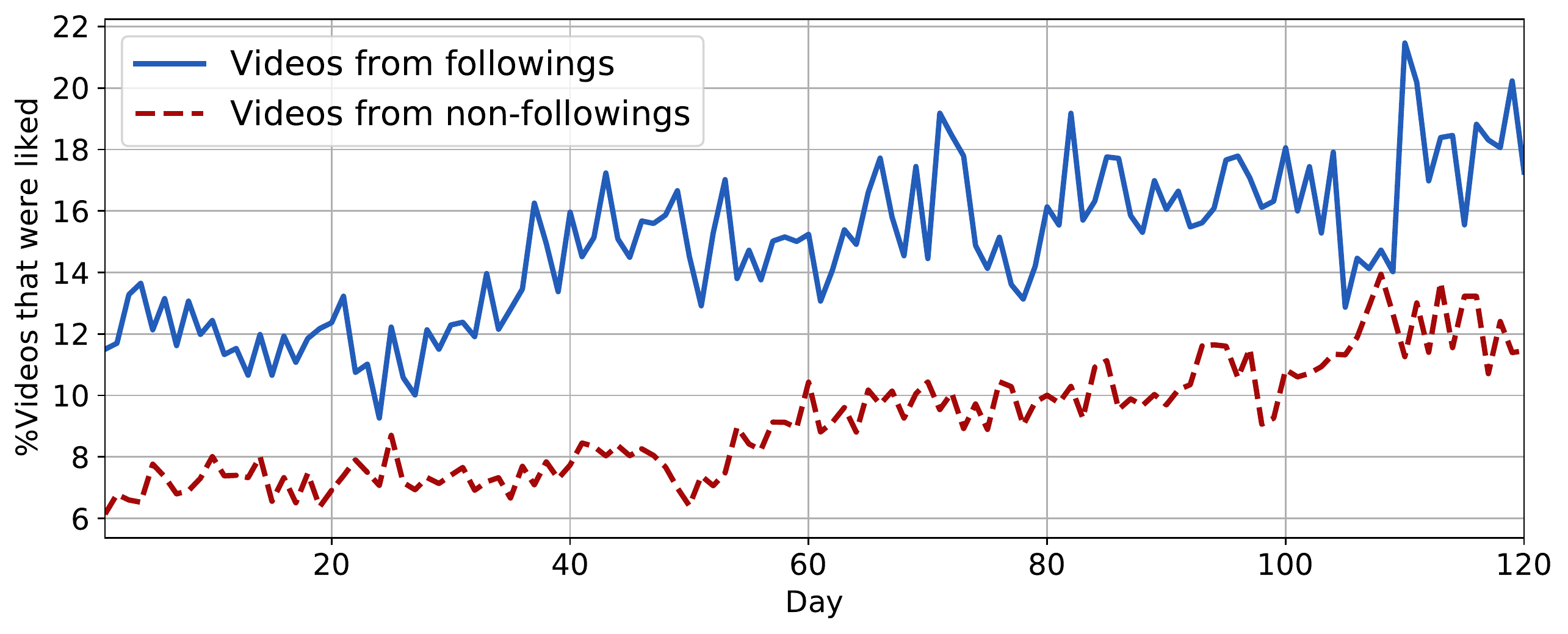}\label{fig:tenure-evolution-engagement}
\caption{Mean interaction (percentage of videos liked) over time. We find that users' engagement using likes increases over the course of the users' tenure on TikTok. }
\label{fig:tenure-evolution-interaction}
\end{figure}

\begin{figure*}[t]
\subfigure[Popularity]{\includegraphics[width=0.33\textwidth]{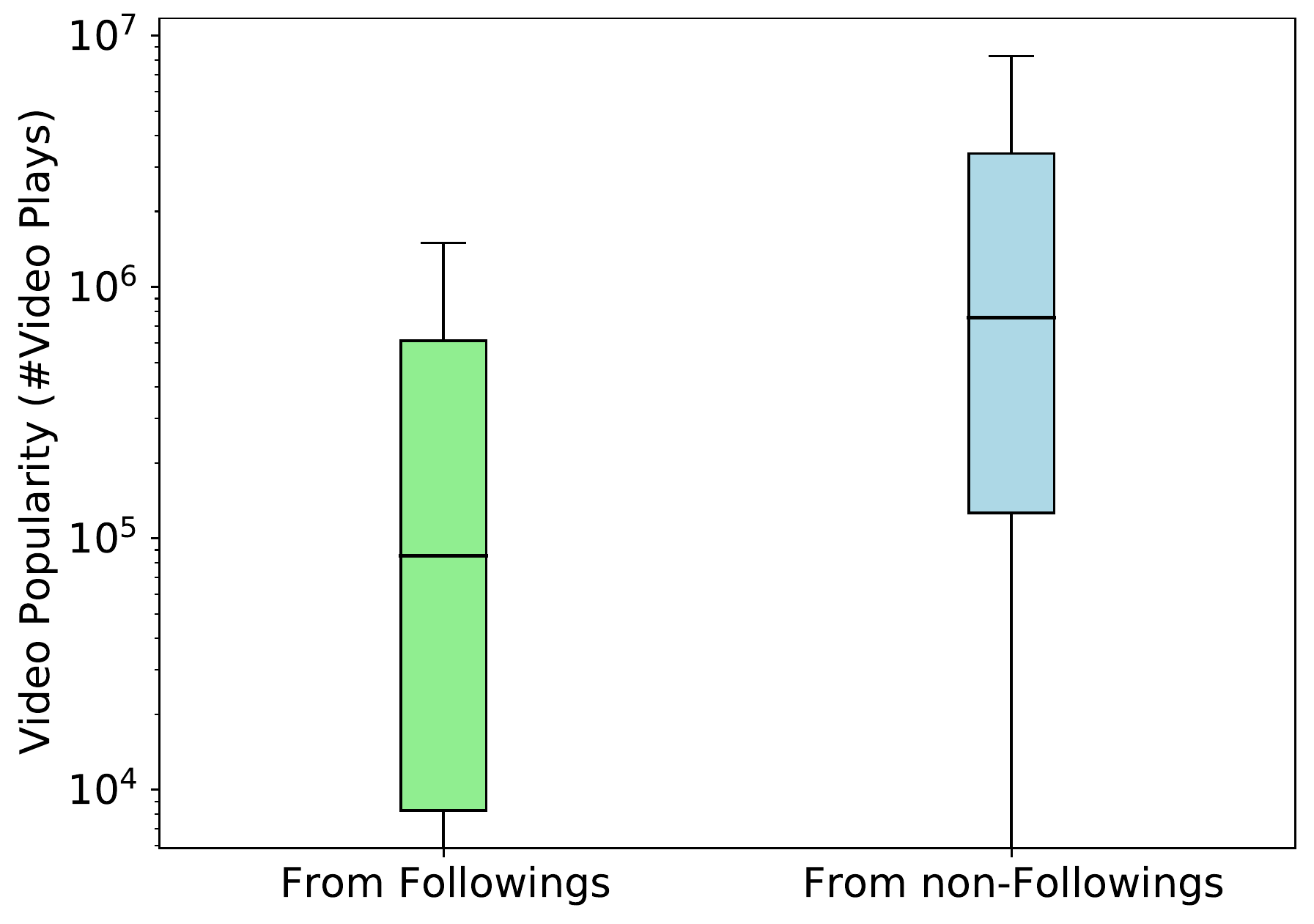}\label{fig:factors-popularity-following}}
\subfigure[Duration]{\includegraphics[width=0.325\textwidth]{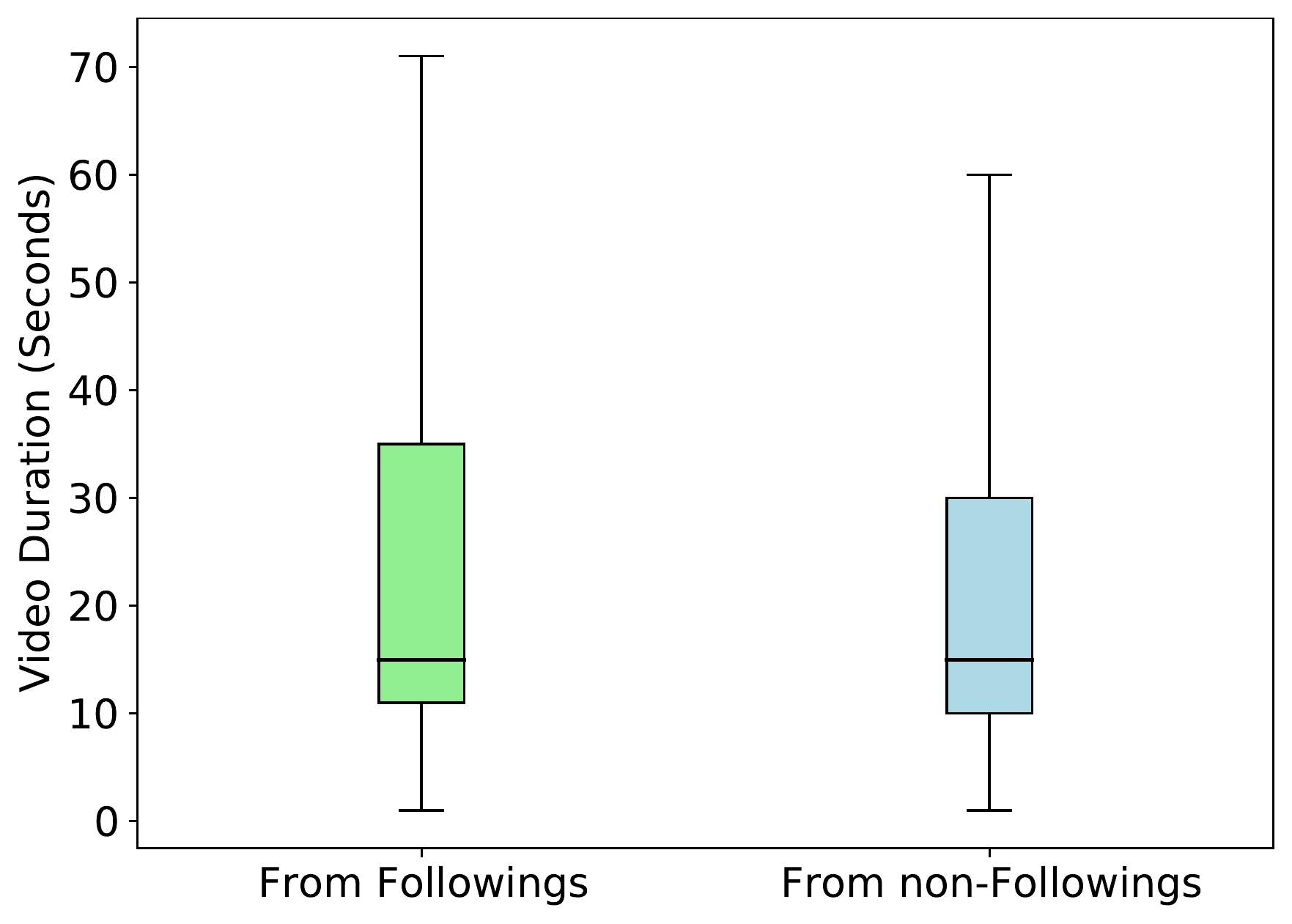}\label{fig:factors-duration-following}}
\subfigure[Age]{\includegraphics[width=0.33\textwidth]{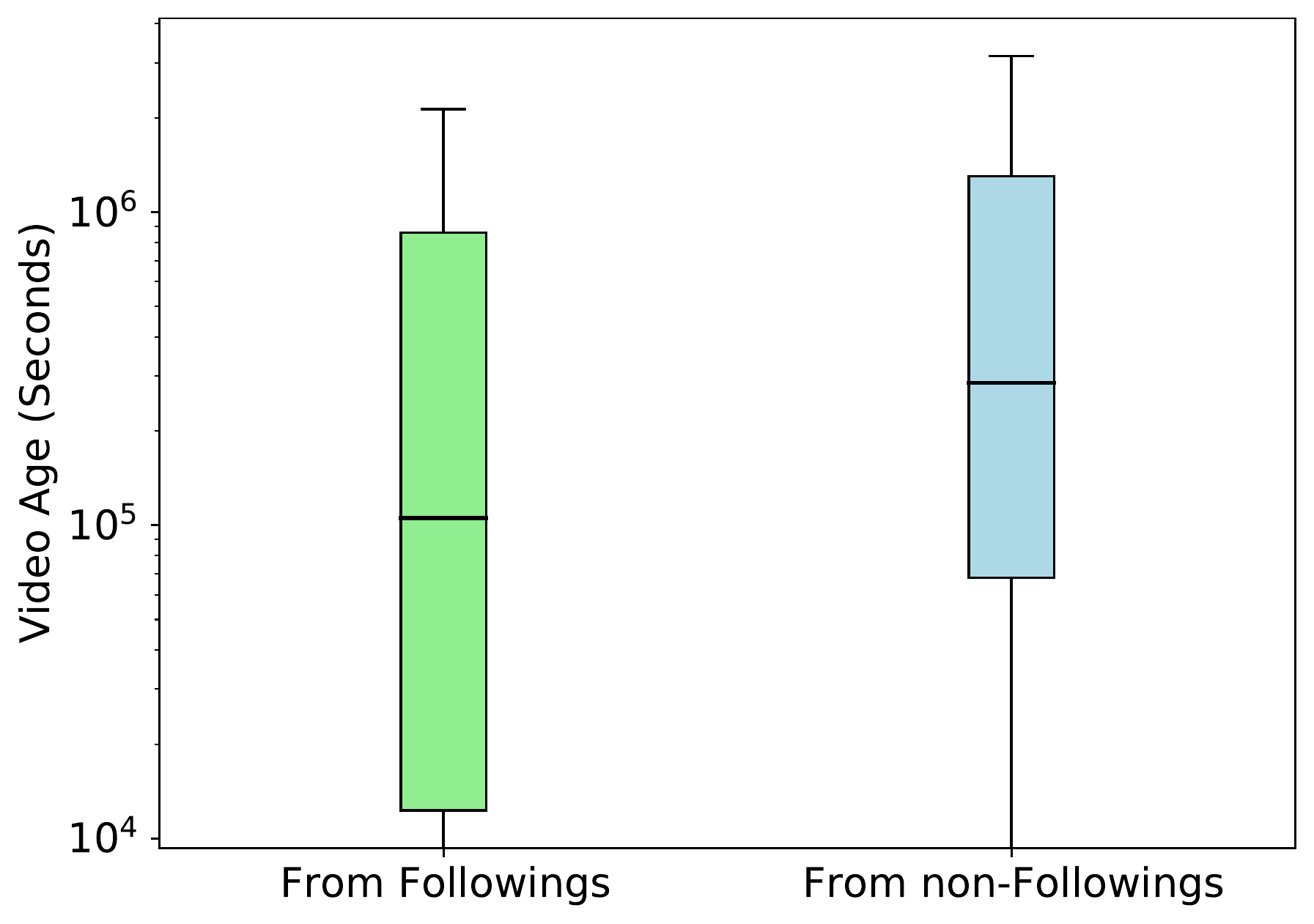}\label{fig:factors-age-following}}
\caption{Comparison of video characteristics for videos from followings and non-followings. Videos from following accounts tend to be less popular, tend to have higher duration, and tend to be newer when compared to videos from non-followings. }
\label{fig:factors_following}
\end{figure*}

\descr{The TikTok participants have an increased video interaction over time and they tend to like more videos from following accounts.}
Next, we look into the interaction signal and how it changes over time. Fig.~\ref{fig:tenure-evolution-interaction} shows how the Interaction signal (i.e., percentage of videos that are liked) changes over time, following the same approach as before.
We observe a substantially different pattern compared to the attention signal; the participants' interactions increase over time for both videos coming from followings and non-followings (from 6\% to 12\% for videos from non-followings, and from 12\% to 18\% for videos from followings, after 120 days), while at the same time, videos from followings receive more interaction compared to videos from non-followings.
The increased interaction on videos from the followings is expected since the participants already expressed explicit interest in those accounts and the following accounts are more likely to be similar to the participants (e.g., be friends), which increases the probability to like their content~\cite{hampton2019you}.

\descr{What are likely reasons why videos from non-following accounts receive more attention?} Previously, we showed that the participants paid less attention to videos from non-followings than from followings, which is counter-intuitive since the participants explicitly expressed interest in that content by following the accounts.
To further understand this phenomenon, here we analyze the videos coming from followings vs. non-followings by comparing various video characteristics. We focus on three video characteristics:
\begin{itemize}
\item \textbf{Video Popularity:} The number of times each video is played on TikTok. We aim to assess whether videos from followings are more popular than videos from non-followings or vice versa.
\item \textbf{Video Duration:} The duration of the video in seconds. We aim to assess whether there are differences regarding the video duration for videos from followings vs. videos from non-followings.
\item \textbf{Video Age:} The difference between the video's creation date and the date that our participant watched the video. We aim to assess whether videos from followings are newer than those from non-following.
\end{itemize}
Fig.~\ref{fig:factors_following} shows the three above-mentioned video characteristics for videos from followings and non-followings.
We observe that videos from non-followings are significantly more popular compared to videos from followings (\emph{median:} 755K vs. \emph{median:} 85K, see Fig.~\ref{fig:factors-popularity-following}), which might be the underlying reason why the participants have increased attention to these videos.
In other words, the participants are more likely to watch until the end videos from non-followings, likely because they are more compelling to users, hence being more popular on TikTok compared to videos from followings.
With respect to the other video characteristics, we find that videos from the followings have slightly higher video duration (\emph{Q1:} 11 vs. \emph{Q1:} 10 and \emph{Q3}: 35 vs.\emph{Q3:} 30, for followings and non-followings, respectively), while for video age we find that videos from followings tend to be newer compared to videos from non-followings (\emph{median}: 3.29 days vs. \emph{median:} 1.21 days for followings and non-followings, respectively).
The differences between the videos from followings and non-followings across all three characteristics are statistically significant, as confirmed by t-tests ($p < 0.05$).
Overall, this analysis highlights that TikTok users are likely to have increased attention to videos from non-followings because they are more popular among TikTok users (as determined by the overall play count on TikTok). This finding is particularly interesting, especially when considering the fact that the participants have a stable consumption of videos from followings over time (see Fig.~\ref{fig:following_views_day}).
Taken altogether, our results might indicate that TikTok's algorithmic recommendations limit the number of videos from following accounts with the goal of maximizing users' attention and retention to the platform.

\descr{Take-Aways.} The main take-away points from our engagement analysis are:
\begin{itemize}
\item A large percentage of videos are not watched till the end (55\%), and most of the participants watched until the end 30\%-50\% of their recommended videos.
\item The attention of the participants is stable over time, and TikTok users tend to watch until the end more videos from non-following accounts compared to videos from following accounts.
\item Videos from non-following accounts are significantly more popular on TikTok compared to videos from following accounts, which likely explains why the participants were more likely to watch them until the end and why TikTok is limiting the recommendation of videos from following accounts (median of 10\% that is stable over time).
\item The user interaction by liking videos has an increasing trend over time; we find a $2\times$ increase after 120 days for videos from followings and a $1.5\times$ increase for videos from non-followings.
\end{itemize}

\section{Discussion \& Conclusion}
In this work, we performed the first empirical analysis of how people engage with short-format videos by undertaking a case study on TikTok.
We implemented a data donation system, recruited 347 users, and collected 9.2M video views.
By analyzing our dataset, we shed some light on how people engage with short-format videos and how their engagement changes over time.
We argue that our study is an important step toward understanding the effectiveness of intelligent information retrieval systems recommending short-format videos to users. 
Below, we discuss our main findings and their implications, the lessons learned by undertaking our data donation, and the limitations of our study.

\subsection{User Engagement on TikTok}

\descr{Increasing number of video views and time spent over time.}
Our analysis shows that the participants spend considerable time daily on TikTok and watch many videos, more so than on other popular video platforms like YouTube~\cite{youtube_use_stats}.
Also, we find that over time, the participants spent an increasing time on the platform, which highlights the need to further explore how addictive the platform is~\cite{meral2021social}, especially when considering that TikTok is extremely popular among people of younger age (e.g., teens).
In particular, we believe that future work should focus on understanding the role of recommendation algorithms in users' experience and potential issues that may arise from recommending short-format videos that the user finds engaging (e.g., addictive behavior).
Overall, we argue that it is important that platforms like TikTok design and offer appropriate well-being nudges that inform users who are spending an increasing amount of time watching videos in an attempt to avoid the development of addiction or other health issues from the continuous exposure of TikTok content (e.g., sleep deprivation).
Note that TikTok, since 2022, has in place the infrastructure and features for such nudges~\cite{tiktok_nudges}. However, users have to explicitly opt-in for these features. 
\revision{We argue that such nudges should be in place by default, assuming that they do not lead to detrimental effects in terms of user autonomy and user engagement. We emphasize the need for further research that aims to analyze and understand the interplay of these nudges and how it affects user autonomy and/or user engagement in short-format video platforms like TikTok.}\revisioncomment{R1}
Overall, we believe that such barriers are even more important on short-format video platforms, which offer an endless stream of recommended short videos since people can watch many videos, and the platform utilizes an algorithm that aims to maximize users' attention and retention.

\descr{Engagement for videos from following accounts.} Our analysis shows that the TikTok participants
watch more videos until the end for videos coming from non-following accounts compared to videos from following accounts (between 44\%-46\% vs. 38\%-42\%).
Looking into the underlying reason for this phenomenon, we find that videos from non-following accounts are, in general, more popular on TikTok, which might be the reason why the participants watched more of such videos until the end (since they are popular and might be more engaging compared to something that they already know since it originates from following accounts).
At the same time, we find that, while the participants follow more accounts over time, the percentage of videos from accounts they follow does not increase over time, which likely indicates that the recommendation algorithm limits the prevalence of the recommended content originating from accounts users follow.
This likely indicates that the recommendation algorithm limits the prevalence of recommendations originating from the users' social networks, possibly in an attempt to further explore users' interests (i.e., recommending videos to explore if the users are interested by receiving implicit feedback like the time spent watching the video).
Finally, when considering user engagement by liking videos, we find that the participants are more likely to like videos from accounts they follow compared to those they do not follow, which is somewhat expected since users usually like content from accounts that might be their friends or they share other similarities with, to send them a direct signal and boost their confidence or self-esteem~\cite{burrow2017many}.

\descr{Video Attention and Algorithmic Recommendations.} \revision{Our results show that only 45\% of the video views are watched until the end, that 70\% of the participants watched until the end between 30\% and 50\% of all the videos in their watch history, and that the participants' attention (i.e., watching until the end) does not increase over time.
There are multiple possible explanations for this phenomenon.
On the one hand, it is likely that the recommendation algorithm tries to maximize users' attention by recommending videos that are likely to be watched until the end, however, the task of predicting attention on short-format videos is too hard or the user signals are too noisy.
Another possible explanation is that the recommendation algorithm can actually do a better job in recommending videos that will be watched until the end but refrain from doing so to increase overall user attention or retention on the platform.
That is, the recommendation algorithm is using ``Reinforcement to increase behavior''~\cite{gazzaniga2010psychological} by mixing positive reinforcement (i.e., videos that are likely to be watched until the end) with some negative reinforcement (i.e., videos that are likely to be skipped by the users).
The inclusion of negative reinforcement is a known phenomenon in Psychology, and it was shown to contribute to addictive behavior in gambling~\cite{weatherly2011testing} or drug use~\cite{koob2013negative}.
Nevertheless, future work is needed to demystify whether the task of recommending short-format videos is hard or whether the recommendation algorithm intentionally includes some negative reinforcement, which can lead to long-term user retention/attention on the platform. }

\subsection{Data Donation Systems and Design Considerations}

Designing a data donation system that uses citizen science to obtain data requires considerations of many factors~\cite{tinati2015designing}. Here, we discuss some of our considerations and provide details on lessons learned from designing and deploying our data donation system. 

\descr{User recruitment.} In this work, our initial goal was to recruit participants exclusively from the U.S., which is why we targeted people from the U.S. on Facebook and explicitly mentioned that we were recruiting U.S. people in our Twitter post.
Despite our initial goal and our ads, our data donation system was publicly available to anyone, and we did not limit access to people coming from specific regions. Hence, we obtained a substantial number of participants from other regions as well (e.g., Africa).
Future work that aims to recruit participants from specific regions or with other restrictions should implement measures that ensure participants meet the study's requirements.
One way to do this is by limiting access to the system to specific IP addresses that come from the region/country of interest.
Despite this, we believe that obtaining global datasets can be extremely useful to the research community.
Given the extensive documentation of bias in academic research toward samples from WEIRD -- western educated industrialized, rich, and democratic -- participant populations~\cite{linxen2021weird}, collecting global datasets can assist in holistically understanding social phenomena through the Web.
Overall, given that few studies in the research community study user behavior in, for instance, Africa, we encourage future work that similarly focuses on a diverse population.

\descr{Malicious users and assessing the quality of donations.}
During our recruitment and data donation procedures, we noticed several instances of malicious users trying to ``trick the system'' to get extra monetary incentives.
We identified three cases of malicious users:
1)~users donating duplicate data (i.e., donating the same data multiple times with different email addresses) to get extra monetary incentives;
2)~users donating their data, then using TikTok for a few more days, and then trying to donate their data again (i.e., duplicate data for the entire period except the few extra days at the end of the data); and
3)~users creating new TikTok accounts, watching only a few videos on TikTok, and then trying to donate their data via our donation system.
Based on these observations, we make the following recommendations.
Researchers implementing data donation systems should design and implement specific countermeasures to detect malicious users who try to donate ``useless'' data.
Concretely, developers of data donation systems should implement features to detect duplicate or near-duplicate donations similar to the ones we implemented.
In addition, to overcome the problem of people creating new accounts for the sake of donating their data, data donation systems can be developed in such a way that they only accept donations that meet a minimum number of days of activity.
For instance, in our case, our donation system rejected all donations where the donation lifespan (i.e., the difference between the last video view and the first video view) was less than three months.
Finally, in our work, we did not notice any attempts to donate completely fabricated data (e.g., randomly computer-generated video URLs and timestamps). However, we argue that this possibility exists, and tech-savvy malicious users may use some techniques in the future depending on how much money they can make from this activity.

\descr{Pricing mechanisms.} Our experience recruiting people to donate their data for research purposes with monetary incentives highlights that those from countries where the incentives offer more purchasing power (e.g., African countries) may be more willing to donate their data. More broadly, an essential aspect of data donation infrastructures is the underlying pricing mechanism since this may affect how willing potential participants will be to donate their data.
For our data donation infrastructure, we set some pre-defined compensation amounts for the viewing video history (\$5), as well as all optional fields (\$1 for each field).
We thought that these prices were reasonable for our purpose and what was asked from the participants. 
However, future work is needed to empirically understand the interplay between data donation and pricing.

\descr{Trustworthiness of the data donation system.} The degree of trust between the participants and the data donation system will likely affect participant recruitment.
We made two design choices in an effort to enhance participant trust. 
First, we provided offline tools that participants can run to anonymize and customize their data before using our data donation system.
Second, we offer users control over which fields to donate.  
Nevertheless, future work on data donation systems should investigate which, if any, of these measures enhance user trust and explore additional measures that may appeal to users, such as stronger formal guarantees that the data donation code is working as described and allowing only the access and queries participants expect to be run on their data.

\descr{Missing data and the need for compliance audits.} By collecting and analyzing user-donated data from TikTok, we noticed some missing data for all the participants (i.e., there was no like data for two months).
The missing data might be due to failures in the logging infrastructure within TikTok or due to lost data. 
Nevertheless, this prompts the need for systematic audits of social media platforms to assess the platforms' compliance with the access rights of data subjects.
Future work can design controlled experiments to assess how accurate and comprehensive the data provided by platforms is, e.g., by using the platform to perform some pre-defined actions, then accessing the provided data and comparing it with the set of pre-defined actions.

\descr{Versality of Data Donation System.} \revision{While in this work, we demonstrated the applicability of data donation systems on TikTok, we argue that the same methodology can be applied to other social media platforms. During this work, we downloaded GDPR data dumps from other platforms (e.g., YouTube Shorts, Instagram, Facebook, Twitter, etc.) to check and assess whether similar data donation methodologies can be applied. We observed that all platforms provide data dumps, with varying degrees of data included, which indicates that performing studies via data donations is a promising avenue for future research. As part of our future work, we aim to expand our data donation system to support data donations from other platforms like YouTube Shorts.}\revisioncomment{R3}

\subsection{Limitations}

We conclude our work with our study's limitations with respect to the data collection and user recruitment, as well as our analysis's limitations.

\descr{Data Collection \& User Recruitment.}
First, our analysis relies on a dataset that comprises a small number of participants (347 TikTok users across the globe). The dataset does not cover a full range of user demographics (e.g., age, gender, geography, see Section~\ref{sec:data_collection}). Due to a lack of public information about the demographics of TikTok's user base, we cannot draw conclusions about the representativeness of the dataset.
Despite this limitation, our results offer a first investigation into how people consume and engage with content on TikTok.
Second, the video metadata collection was done post-hoc, hence we are unable to obtain a holistic view of all videos referenced in the data since approximately 17\% of all videos were not accessible during the video metadata collection period.
Third, the dataset focuses on a single platform (TikTok), hence we are unable to make comparisons with other platforms.
Obtaining behavioral traces across multiple platforms is challenging due to diverse features on each platform and differences in the data provided using data access requests across platforms (e.g., YouTube not providing the like data).
Finally, the recruitment methodology of the dataset might introduce some biases as it is likely that recruited participants are, in general, more willing to donate their behavioral traces for research purposes than other TikTok users.
Additionally, given the extensive documentation of bias in academic research toward samples from WEIRD -- western educated industrialized, rich, and democratic -- participant populations (see, e.g., the study from Linxen et al.~\cite{linxen2021weird}), we analyzed a dataset that includes participants across the globe. 
Given that few studies in academia measure social media user engagement in, e.g., Africa,  we encourage future work that similarly focuses on a diverse user population, which can serve to evaluate the representativeness of the dataset. 

\descr{Analysis.} Our analysis has some limitations that are worth mentioning.
\revision{First, our participants started using TikTok at different periods, and while we aligned the viewing histories in our analysis, we can not assess whether the collected dataset represents a stable state for each participant. That is, some participants may have just started using the platform and be excited about using a new platform.}
\revision{Second, our engagement analysis focuses on the participants' liking behavior. While commenting and sharing videos is a crucial engagement signal, we focused on likes as the overwhelming majority of the user engagement is done via likes in our dataset, and because of the fact that we do not have information about the videos, comments were made on.}\revisioncomment{R1}
Third, our analysis of user attention (i.e., how many videos are watched till the end) does not consider a potential ceiling effect (i.e., users having limited time to view videos). A ceiling effect will lead to users raising their bar on what to watch till the end, leading to the stagnation of the attention metric.
Also, our work does not analyze user retention on the platform or users who drop from the platform (i.e., users who stop using the platform). We believe that this is a fascinating avenue for future work; we do not perform this analysis in this work, as it will require a new data collection and recruitment methodology. This is because, when recruiting TikTok users, we required them to be active on TikTok for at least the last three months. Therefore, our dataset only includes data for active users on TikTok, and we do not have any data about lost users.
Additionally, our analysis makes some assumptions about the direct relationship and influence of the recommendation algorithm on user engagement. It is possible that there are other confounding factors that might affect user engagement beyond the recommendation algorithm and are not accounted for in our work.
Another important limitation is that even though the data obtained via donations are quite detailed, some of our results are based on inferences, such as the time that each participant spent watching videos.
Due to this, we are unable to accurately infer the viewing durations for video views that are above our determined threshold following the methodology by~\cite{halfaker2015user}.
Therefore, our results based on the time spent on the platform should be considered lower-bound results, as we exclude all the video views at the end of each inferred session.
To conclude, many of our limitations reflect fundamental challenges with data donation and data analysis on data donated by real users directly to study and audit closed platforms externally.

\bibliographystyle{ACM-Reference-Format}

\appendix

\section{Comparative Analysis across Regions}

\begin{figure}[t]
\subfigure[Volume]{\includegraphics[width=\columnwidth]{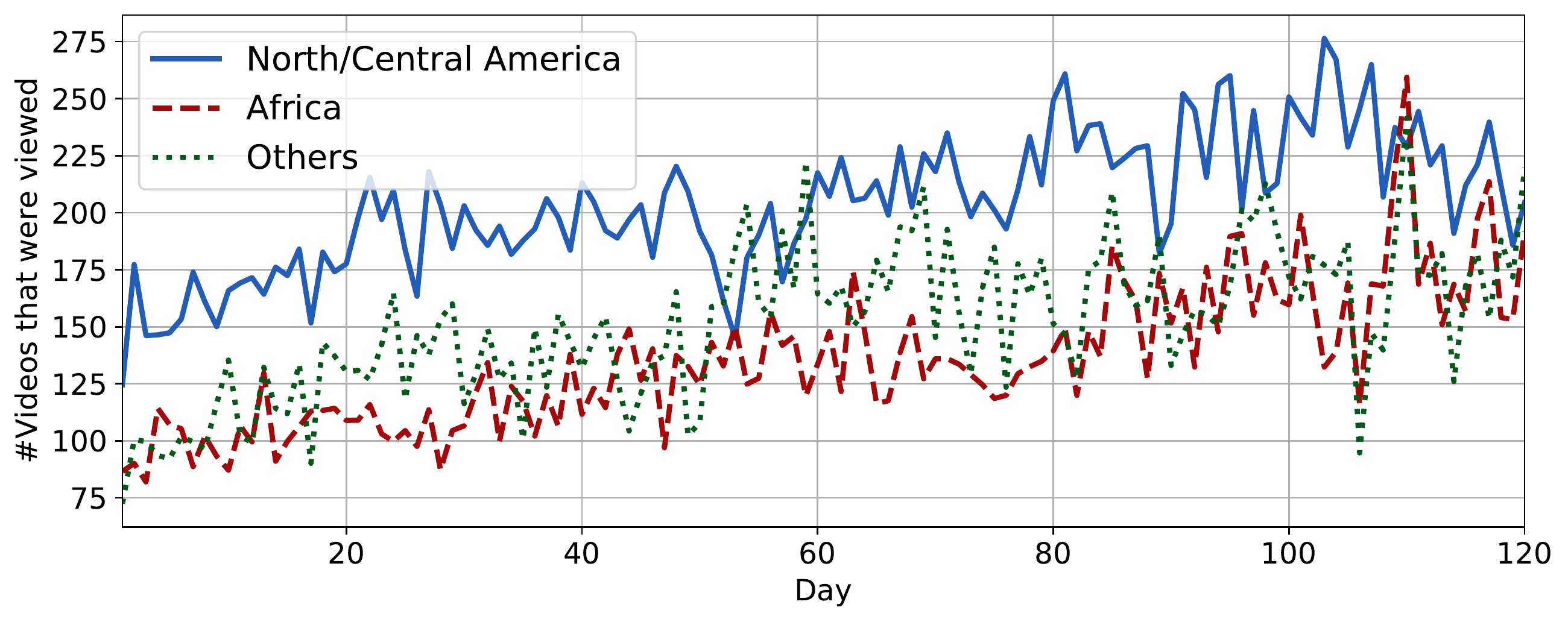}\label{fig:tenure-evolution-views-region}}
\subfigure[Time]{\includegraphics[width=\columnwidth]{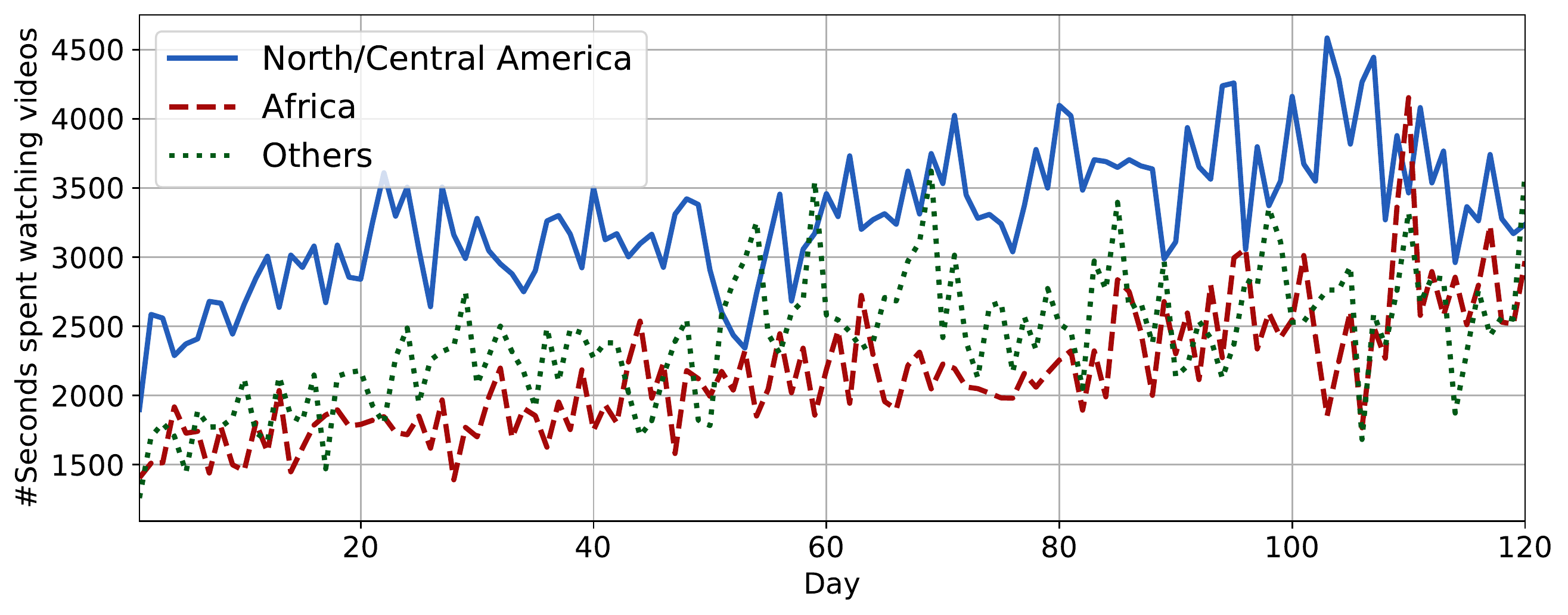}\label{fig:tenure-evolution-time-region}}

\caption{\revision{Changes in the volume of watched videos and time spent on TikTok over time for participants across regions. We observe that over time, for all geographical regions, there is an increase in the volume of videos watched per day and the daily time spent on TikTok.}\revisioncomment{R4}}
\label{fig:region-comparative}
\end{figure}

\revision{Our dataset is quite diverse regarding the users' geographical locations, with most users being from North/Central America and Africa.
Here, we undertake a comparative analysis of user behavior and engagement across the various geographical locations in our dataset.
Fig.~\ref{fig:region-comparative} shows the number of videos watched and time spent per day by participants across multiple regions in our dataset. 
We observe that the general trend of increasing number of videos and time spent daily over time holds for participants, irrespective of their geographical region.
Also, we note that in our dataset, participants from North/Central America tend to watch more videos daily and spend more time on TikTok than participants from Africa or other geographical regions.
}

\begin{figure}[t]
\includegraphics[width=\columnwidth]{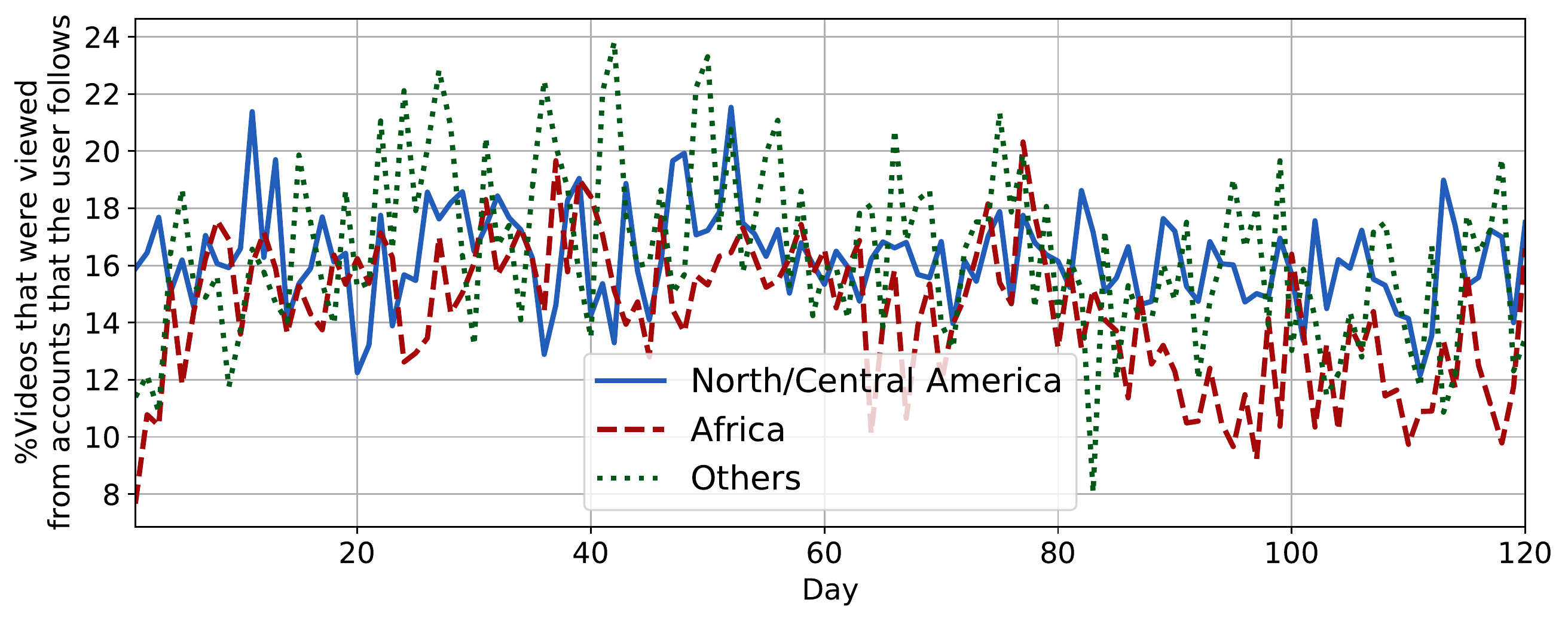}
\caption{Average number of video views from users that our participants already followed for participants across geographical regions.}
\label{fig:following-regions}
\end{figure}

\revision{With respect to the number of video views from accounts that our participants follow, we find no substantial differences across the demographic regions. Specifically, by considering all the video views in our dataset, we find that participants from North/Central America watched 10.91\% of their videos from accounts they already followed, while for participants from Africa and Other regions, we find a percentage of 10.72\% and 8.28\%, respectively.
Also, by looking into how the number of videos from following accounts changes over time for participants across geographical regions (see Fig.~\ref{fig:following-regions}), we find no substantial differences across geographical regions. Overall, these results show that the recommendation algorithm recommends a similar proportion of videos from accounts that a user already follows, irrespective of a user's geographical location.}

\begin{figure}[t]
\subfigure[Attention]{\includegraphics[width=\columnwidth]{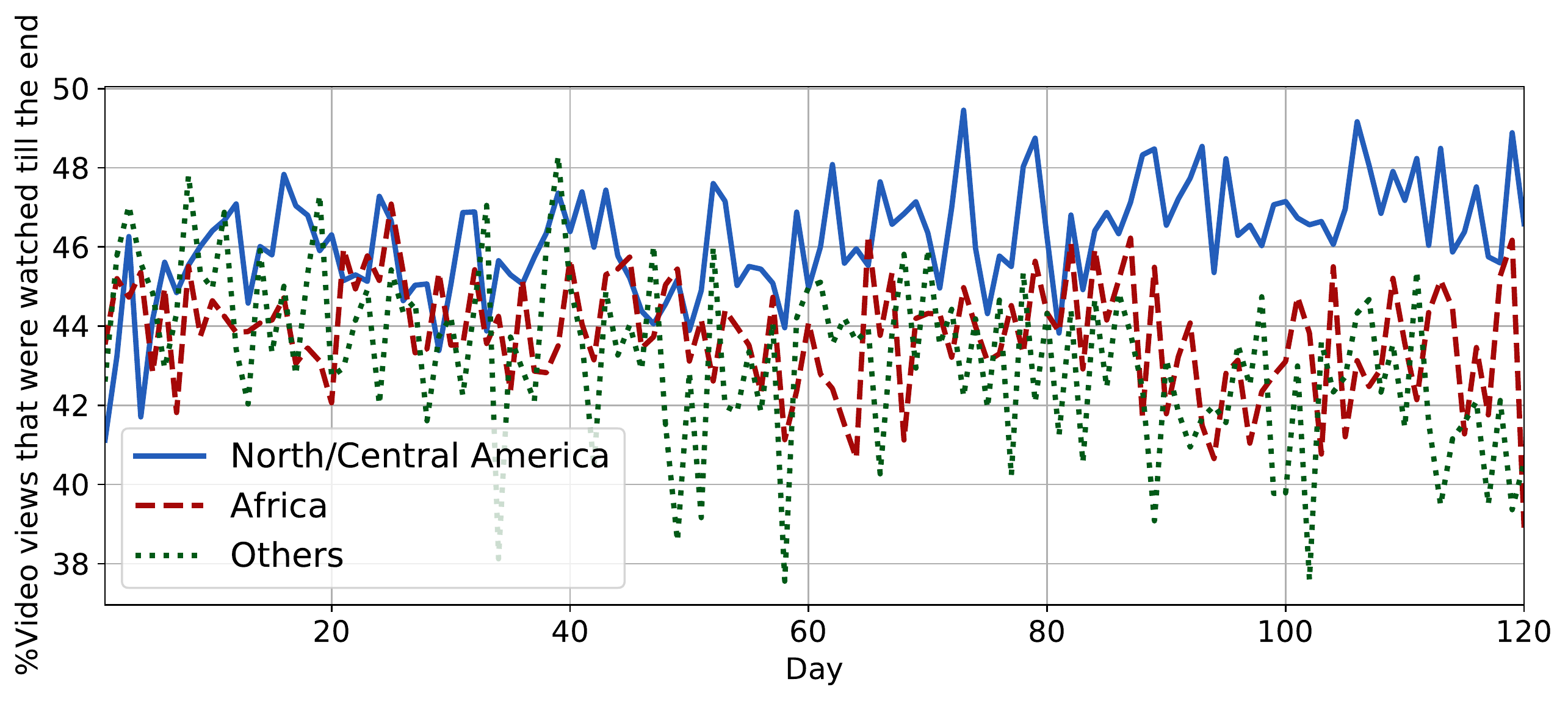}\label{fig:tenure-watched-region}}
\subfigure[Likes]{\includegraphics[width=\columnwidth]{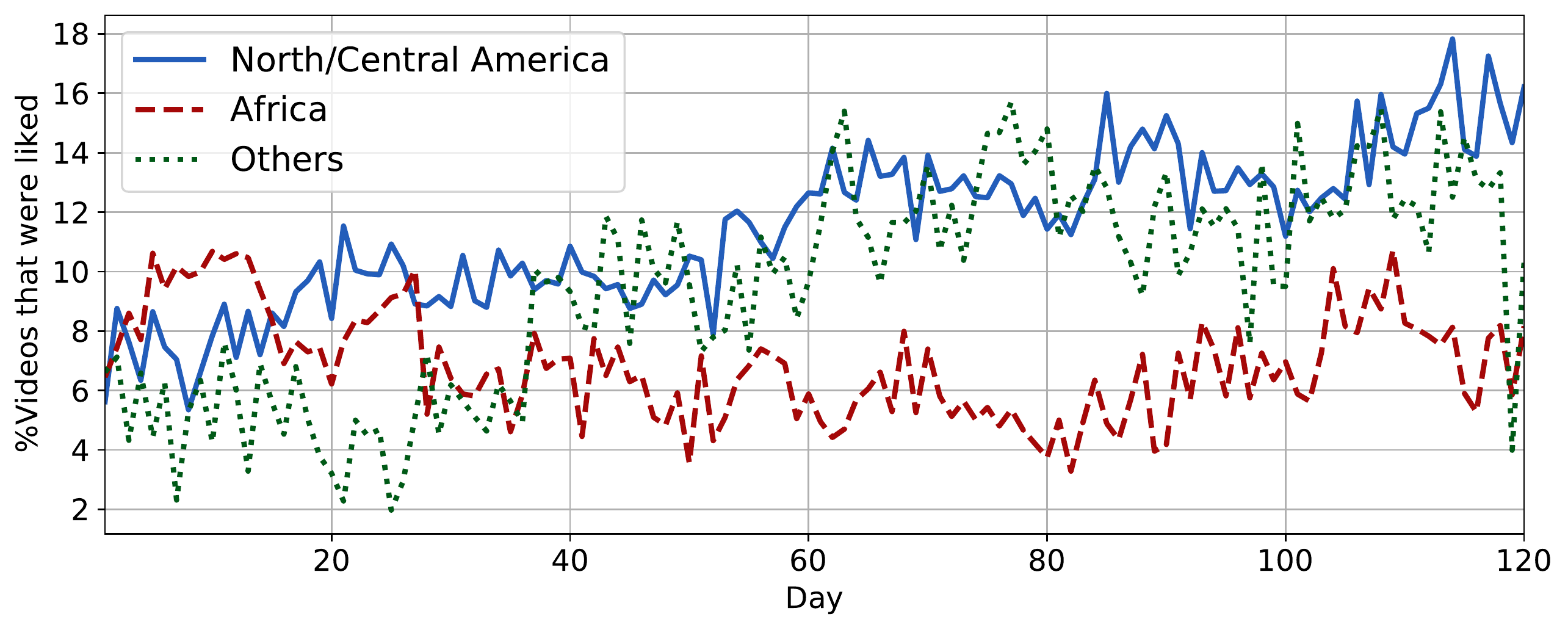}\label{fig:tenure-evolution-time-region}}

\caption{\revision{Changes in the Attention metric (mean number of watched till the end videos) and engagement through liking for participants across various geographical regions. We report average values for participants grouped based on their geographical regions.}\revisioncomment{R4}}
\label{fig:region-comparative-engagement}
\end{figure}

\revision{We also plot the changes over time in terms of the Attention metric and the percentage of videos that are liked by participants across the various geographical regions in Fig.~\ref{fig:region-comparative-engagement}.
Similarly to the aggregate results (for all participants), we observe that the Attention metric is stable over time, indicating that irrespective of the participants' geographical region, participants watch a similar percentage of videos till the end (see Fig.~\ref{fig:tenure-watched-region}).
Concerning the participants' engagement through liking videos, we observe some differences across the participants based on their geographical regions. For participants from North/Central America, we observe a substantial increase in the percentage of liked videos, from 6\% to 16\% after 120 days. 
On the other hand, for participants from Africa, we observe that the engagement via liking videos increases for the first 15 days (from 6\% to 10\%), and then the liking behavior becomes more stable with participants liking between 6\% and 8\% of videos per day).}

\revision{Taken altogether, this comparative analysis shows that, based on our dataset, the participants' activity, with the exception of liking behavior over time for participants in Africa, does not exhibit substantial differences.}
\revisioncomment{R4}

\end{document}